\documentclass[12pt]{article}

\newlength{\abstractwidth}
\usepackage{amsmath}
\usepackage{amsfonts}
\usepackage{amssymb}

\usepackage{epsf}
\usepackage{color}
\usepackage{graphicx}
\usepackage{tikz}
\usepackage{dsfont}
\usepackage{subfigure}

\usepackage{hyperref}
\definecolor{darkred}{rgb}{0.8,0.1,0.1}
\hypersetup{colorlinks=true, linkcolor=darkred, citecolor=blue, linktoc=page}

\usetikzlibrary{arrows,shapes,positioning}
\usetikzlibrary{decorations.markings}
\usepackage[rightcaption]{sidecap}
\tikzstyle arrowstyle=[scale=1]
\tikzstyle directed=[postaction={decorate,decoration={markings,
    mark=at position .65 with {\arrow[arrowstyle]{stealth}}}}]
\tikzstyle reverse directed=[postaction={decorate,decoration={markings,
    mark=at position .65 with {\arrowreversed[arrowstyle]{stealth};}}}]
\usetikzlibrary{positioning}

\usepackage{amssymb}
\usepackage{latexsym}

\renewcommand{\thefootnote}{\fnsymbol{footnote}}
\renewcommand{\thanks}[1]{\footnote{#1}}
\newcommand{\starttext}{
\setcounter{footnote}{0}
\renewcommand{\thefootnote}{\arabic{footnote}}}

\newcommand{\bea}{\begin{eqnarray}}
\newcommand{\eea}{\end{eqnarray}}
\newcommand{\be}{\begin{eqnarray}}
\newcommand{\ee}{\end{eqnarray}}

\def\cA{{\cal A}}
\def\cB{{\cal B}}
\def\cC{{\cal C}}

\def\cF{{\cal F}}
\def\cG{{\cal G}}
\def\cH{{\cal H}}
\def\cI{{\cal I}}

\def\cS{{\cal S}}

\def\cY{{\cal Y}}

\def\ZZ{{\mathbb Z}}
\def\RR{{\mathbb R}}

\def\CC{{\mathbb C}}

\def\Re{{\rm Re \,}}
\def\Im{{\rm Im \,}}

\def\det{{\rm det \,}}

\def\half{{1\over 2}}
\def\p{\partial}

\def\eps{\epsilon}

\def\ep{\varepsilon}

\def\no{\nonumber}
\def\sm{\smallskip}

\def\pq{$(p,q)$}
\def\pqseven{$[p,q]$}

\makeatletter\def\l@subsubsection#1#2{}%
\makeatother

\begin{document}
\starttext
\setcounter{footnote}{0}

\begin{flushright}
2017 May 25
\end{flushright}

\vskip 0.3in

\begin{center}

{\Large \bf Warped $AdS_6\times S^2$ in Type IIB supergravity III}

\vskip 0.2in

{\large \bf Global solutions with seven-branes}

\vskip 0.5in

{\large   Eric D'Hoker$^{a,b}$,  Michael Gutperle$^{a}$ and Christoph F.~Uhlemann$^{a}$} 

\vskip 0.15in

{ \sl 
${}^a$ Mani L. Bhaumik Institute for Theoretical Physics}\\
{\sl  Department of Physics and Astronomy}\\
{\sl University of California, Los Angeles, CA 90095, USA}

\vskip 0.2in

{\sl ${}^b$ Kavli Institute for Theoretical Physics}\\
{\sl University of California Santa Barbara, CA 93106, USA}

\vskip 0.2in

{\tt \small dhoker@physics.ucla.edu; gutperle@ucla.edu; uhlemann@physics.ucla.edu}

\vskip 0.5in

\begin{abstract}

We extend our previous construction of global solutions to Type IIB supergravity  that are invariant under the  superalgebra $F(4)$ and are realized on a spacetime of the form $AdS_6 \times S^2$ warped over a Riemann  surface $\Sigma$ by allowing the supergravity fields to have non-trivial $SL(2,\RR)$ monodromy at isolated punctures on $\Sigma$. We obtain explicit solutions for the case where $\Sigma$ is a disc, and the monodromy generators are parabolic elements of $SL(2,\RR)$ physically corresponding to the monodromy allowed  in  Type IIB string theory. On the boundary of $\Sigma$ the solutions exhibit singularities at isolated points which correspond to semi-infinite five-branes, as is familiar from the global solutions without monodromy. In the interior of $\Sigma$, the solutions are everywhere regular, except at the punctures where $SL(2,\RR)$ monodromy resides and which physically correspond to the locations of \pqseven{}  seven-branes. The solutions have a compelling physical interpretation corresponding to fully localized five-brane intersections with additional seven-branes, and provide candidate holographic duals to the five-dimensional superconformal field theories realized on such intersections.

\end{abstract}
\end{center}

\baselineskip=15pt
\setcounter{equation}{0}
\setcounter{footnote}{0}

\newpage
\tableofcontents

\newpage

\section{Introduction}
\setcounter{equation}{0}
\label{sec:intro}

Five-dimensional superconformal field theories (SCFTs) exhibit many intriguing and exotic properties, 
including the uniqueness of the exceptional superconformal symmetry algebra $F(4)$,  the possibility for exceptional global symmetries, the absence of a useful Lagrangian description, and many non-trivial dualities and relations to theories in other dimensions.

\sm

In the absence of a conventional Lagrangian description, the theories have been accessed indirectly, for example as non-trivial UV fixed-points of five-dimensional gauge theories considered on the Coulomb branch  or as low-energy description of certain brane configurations in string theory or M-theory on Calabi-Yau manifolds \cite{Seiberg:1996bd,Intriligator:1997pq}.
A  very fruitful approach has been to engineer these theories using brane constructions in Type IIB string theory.
Five-dimensional gauge theories can be realized on the world-volume of D5-branes that are  suspended between semi-infinite external \pq{} 5-branes \cite{Aharony:1997ju,Aharony:1997bh}. In the limit where these brane webs collapse to a fully localized intersections of \pq{} 5-branes one recovers the SCFTs at the origin of their moduli spaces.
While the string theory constructions provide access to many features of the 5d SCFTs and have led to many insights, the corresponding supergravity solutions in Type IIB supergravity are the prerequisite for utilizing AdS/CFT as tool for comprehensive quantitative analyses.
In recent work we have constructed large classes of warped $AdS_6$ solutions in Type IIB supergravity that are in direct correspondence with fully localized 5-brane intersections in Type IIB string theory \cite{D'Hoker:2016rdq,DHoker:2016ysh,DHoker:2017mds}. They allow for quantitative analyses of the theories realized on intersections of 5-branes, including as a first step the study of  free energies and entanglement entropies \cite{Gutperle:2017tjo}.

\sm

The 5-brane web constructions in Type IIB string theory can be generalized considerably by including 7-branes \cite{DeWolfe:1999hj,Yamada:1999xr}.
External 5-branes are allowed to terminate on 7-branes and 7-branes may be added into the open faces of the web. The Hanany-Witten brane creation effect \cite{Hanany:1996ie} provides a way to relate  certain webs with 7-branes to webs without 7-branes.
Many recent insights are based on manipulations involving 7-branes, including new dualities between 5d theories from branch cut moves, the construction of gauge theory descriptions for 5d uplifts of 4d class S theories, the realization of theories violating the flavor bounds of \cite{Seiberg:1996bd,Intriligator:1997pq}\footnote{Recent attempts towards a more complete classification can be found in \cite{Xie:2017pfl,Jefferson:2017ahm}.} and connections to 6d SCFTs \cite{Benini:2009gi,Taki:2014pba,Bergman:2014kza,Kim:2015jba,Hayashi:2015fsa,Hayashi:2015zka}.
These observations provide a clear motivation for the construction of warped $AdS_6$ solutions in Type IIB supergravity corresponding to 5-brane intersections which include 7-branes.
 7-branes placed inside the faces of a 5-brane web, for example, should be directly accessible via supergravity solutions corresponding to the conformal limit of the web. 
In the present paper we will construct warped $AdS_6$ solutions to Type IIB supergravity  which include 7-branes.

\sm

The geometry of the solutions constructed in \cite{D'Hoker:2016rdq,DHoker:2016ysh,DHoker:2017mds} takes the form of $AdS_6\times S^2$ warped over a two-dimensional Riemann surface $\Sigma$ with boundary, and the solutions realize holographically the unique $F(4)$ superconformal algebra in 5d \cite{Kac:1977em,Shnider:1988wh}. 
The solutions are specified in terms of two locally holomorphic functions $\cA_\pm$ on $\Sigma$ and a crucial feature is that the differentials of these functions have poles on the boundary of $\Sigma$. At these poles the geometry approaches that of \pq{} 5-branes with the charges given by the residues, and this allows for a clear mapping between 5-brane intersections and supergravity solutions. For the global solutions constructed explicitly so far $\Sigma$ was taken to be a disc \cite{DHoker:2016ysh,DHoker:2017mds}.
These solutions will provide the basis for the construction of solutions with 7-branes.

\sm

The distinct feature of 7-branes, amongst the brane solutions in Type IIB supergravity,  is the defect they create in the space transverse to their world-volume, and the non-trivial monodromy  the axion and dilaton fields exhibit around this defect. The duality group of Type IIB supergravity is $SL(2,\RR)$ and, mathematically, the axion-dilaton field  and the three-form field strengths might have monodromy with arbitrary values in $SL(2,\RR)$. 
Physically, however, we are interested in supergravity solutions which embed into Type IIB string theory.  The duality group of Type IIB string theory is $SL(2,\ZZ)$, and string  theory solutions only allow for $SL(2,\ZZ)$-valued monodromy. For example the monodromy of a D7-brane  leaves the dilaton invariant and shifts the axion field by 1, corresponding to a parabolic element of $SL(2,\ZZ)$. Just as strings and 5-branes, 7-branes transform non-trivially under $SL(2,\ZZ)$ so that a general  7-brane carries a charge labeled by a pair of integers \pqseven{} which  specify the monodromy around the 7-brane \cite{Gaberdiel:1997ud,Gaberdiel:1998mv,DeWolfe:1998eu}. In supergravity, $SL(2,\ZZ)$ is replaced by $SL(2,\RR)$,  $p$ and $q$ become real numbers, and the monodromy can be a generic parabolic element of $SL(2,\RR)$.

\sm

The supersymmetry conditions on branes allow for the preservation of the full $F(4)$ superalgebra in the presence of both 5-branes and 7-branes, and we shall henceforth restrict to solutions with this full symmetry. 
Preserving the full  $F(4)$ requires the 7-branes to be located at isolated points or punctures  in the interior of the surface $\Sigma$, around which the supergravity fields  have non-trivial monodromy given by a parabolic element of $SL(2,\RR)$. 
We will show that  the monodromy of the supergravity fields around the punctures can be realized by suitable monodromies of the locally holomorphic functions $\cA_\pm$ which parametrize the solutions, and we will explicitly construct such $\cA_\pm$ and the corresponding supergravity solutions. 
We will allow for an arbitrary number of punctures with mutually commuting monodromies. These are the appropriate monodromies for an arbitrary number of mutually local 7-branes, and we will show that the asymptotic behavior of the solutions near the punctures indeed approaches the  form expected on physical grounds.

\sm

The remainder of the paper is organized as follows. In sec.~\ref{sec:review} we will review the global solutions constructed in \cite{D'Hoker:2016rdq,DHoker:2016ysh,DHoker:2017mds} and highlight the points that will be relevant for the construction of solutions with 7-brane monodromy.
The actual construction will be carried out in sec.~\ref{sec:sol}, where we explicitly set up the holomorphic data for solutions with monodromy and derive the regularity conditions constraining the parameters.
We will also show that the supergravity fields close to the punctures match to the expected form for \pqseven{} 7-branes.
In sec.~\ref{sec:examples} we will solve the regularity conditions and present explicit example solutions, showing that the solutions indeed have the desired properties.
The connection to 5-brane webs with 7-branes will be discussed in more detail in sec.~\ref{sec:5-brane-webs} and we close with a discussion in sec.~\ref{sec:discussion}.

\newpage

\section{Review of solutions without monodromy}
\setcounter{equation}{0}
\label{sec:review}

In this section we will briefly review the local solutions to Type IIB supergravity with 16 supersymmetries and metric of the form $AdS_6\times S^2$ warped over a Riemann surface $\Sigma$ as constructed in \cite{D'Hoker:2016rdq}, the regularity conditions they have to satisfy and the global solutions without monodromy constructed in \cite{DHoker:2017mds}. The global solutions without monodromy will be the starting point for the construction of solutions with monodromy in the next section.

\subsection{Supergravity fields in terms of holomorphic data}

The general local solution with 16 supersymmetries and $SO(2,5)\times SO(3)$ isometry can be expressed in terms of two locally holomorphic functions $\cA_\pm$ defined on the Riemann surface $\Sigma$ with so far arbitrary topology.
The symmetry requirement restricts the metric and two-form field strength to take the form
\bea
\label{2a1}
ds^2 & =  & f_6^2 \, ds^2 _{AdS_6} + f_2 ^2 \, ds^2 _{S^2} + 4\rho^2 | dw |^2
\no \\
  F_{(3)} & = & d \cC  \wedge {\rm vol}_{S^2}
\eea
with $f_6^2$, $f_2^2$ $\rho^2$ real functions on $\Sigma$ while $\cC$ is an in general complex function on $\Sigma$.
The four-form field vanishes.
The functions appearing in the ansatz can be conveniently expressed in terms of $\cA_\pm$ by using the composite objects
\begin{align}
 \kappa ^2 & =  - |\p_w \cA_+|^2 + |\p_w \cA_-|^2
 &
 \cG & =  |\cA_+|^2 - |\cA_-|^2 + \cB + \bar \cB
 \nonumber\\
 \p_w \cB &= \cA_+ \p_w \cA_- - \cA_- \p_w \cA_+
 & 
 R+\frac{1}{R} &=  2+ { 6 \, \kappa^2 \, \cG \over |\partial_w\cG|^2}
 \label{eq:comp}
\end{align}
where $\cB$ is defined up to an integration constant.
The metric functions then take the form
\begin{align}\label{eq:f2f6rho}
f_6^2&=c_6^2 \sqrt{6\cG} \left ( \frac{1+R}{1-R} \right ) ^{1/2}
&
f_2^2&=\frac{c_6^2}{9}\sqrt{6\cG} \left ( \frac{1-R}{1+R} \right ) ^{3/2}
&
\rho^2={ \kappa^2 \over \sqrt{6\cG} } \left (\frac{1+R}{1-R} \right ) ^{1/2}
\end{align}
The remaining fields are the axion-dilaton scalar $B$, which is given by
\bea 
\label{2a6}
B ={\p_w \cA_+ \,  \partial_{\bar w} \cG - R \, \p_{\bar w} \bar \cA_-   \partial_w \cG \over 
R \, \p_{\bar w}  \bar \cA_+ \partial_w \cG - \p_w \cA_- \partial_{\bar w}  \cG}
\eea
and the complex function $\cC$ parametrizing the two-form gauge field, which reads
\bea
\label{2a7}
 \cC = \frac{4 i c_6^2}{9}\left (  
{\p_{\bar w} \bar \cA_- \, \p_w \cG \over \kappa ^2} 
- 2 R \, {  \partial_w \cG \, \p_{\bar w} \bar \cA_- +  \partial_{\bar w}  \cG \, \p_w \cA_+ \over (R+1)^2 \, \kappa^2 }  
 - \bar  \cA_- - 2 \cA_+ \right )
\eea

\subsection{Regularity conditions and global solutions}\label{sec:global-reg}

For physically sensible solutions additional regularity conditions are required, and these can be 
expressed concisely as conditions on the composite quantities $\kappa^2$ and $\cG$ defined in (\ref{eq:comp}). 
To have the metric functions $f_6^2$, $f_2^2$, $\rho^2$ positive in the interior of $\Sigma$ it is sufficient to require
\bea
\label{2b1}
\kappa ^2  >  0
\hskip 1in
\cG  >  0
\eea
A smooth and geodesically complete ten-dimensional geometry  can be realized by shrinking the $S^2$ on the boundary $\partial\Sigma$ of $\Sigma$,  which amounts to the additional conditions
\bea
\label{2b2}
\kappa ^2 \Big |_{ \p \Sigma} =0 \hskip 1in  \cG \Big |_{\p \Sigma} =0
\eea
This finishes the general discussion of the regularity conditions. 
As shown in \cite{DHoker:2017mds} they can be satisfied by choosing $\Sigma$ to be the upper half plane and
the locally holomorphic functions as 
\be\label{eq:A0}
 \cA_\pm (w) =\cA_\pm^0+\sum_{\ell=1}^L Z_\pm^\ell \ln(w-p_\ell)
\ee
where $p_\ell$ for $\ell=1,\cdots,L$ denote the $L$ poles of the differentials $\partial_w\cA_\pm$; they lie on the real line which is the boundary $\p \Sigma$.
The residues of $\p_w \cA_\pm$ at these poles, $Z_\pm^\ell$, are expressed in terms of $L-2$ zeros $s_n$, $n=1,\cdots,L-2$ in the upper half plane, with the restriction that at least one of them must be in the interior of $\Sigma$. They take the form
\be
Z_+^\ell  =
 \sigma\prod_{n=1}^{L-2}(p_\ell-s_n)\prod_{k \neq\ell}^L\frac{1}{p_\ell-p_k}
\ee
with an overall complex normalization parametrized by $\sigma$, and $Z^\ell_- = - \overline{Z_+^\ell}$.
The locally holomorphic functions constructed this way satisfy the regularity conditions on $\kappa^2$, produce $\cG$ constant along each boundary component free of poles and $\cG>0$ in the interior of $\Sigma$ if $\cG=0$ on the boundary. The only condition left to satisfy therefore is $\cG=0$ on the boundary, which constrains the parameters to satisfy
\be
 \cA^0 Z_-^k + \bar \cA^0 Z_+^k 
+ \sum _{\ell \not= k }Z^{[\ell k]} \ln |p_\ell - p_k| =0
\ee
for $k=1,\cdots,L$, where we have defined  $2\cA^0=\cA_+^0-\bar \cA_-^0$ and $Z^{[\ell k]}=Z_+^\ell Z_-^k-Z_+^kZ_-^\ell$.
Regularity of the string-frame geometry near the poles furthermore requires $c_6^2=1$.

\subsection{\texorpdfstring{$SU(1,1)$}{SU(1,1)} transformations}

The $SL(2,\mathbb{R})\sim SU(1,1)$ duality symmetry transformations of Type IIB supergravity have been realized on the locally holomorphic data $\cA_\pm$ and on the composite quantities $\kappa ^2$ and $\cG$ in \cite{D'Hoker:2016rdq}.
Parametrizing a generic $SU(1,1)$ transformation by $u,v\in \CC$ with $|u|^2-|v|^2=1$,
the locally holomorphic functions $\cA_\pm$ transform as follows
\bea
\label{Atrans}
\cA_+ & \to & \cA_+'=  + u \cA_+ - v \cA_- + a_+ 
\no \\
\cA_- & \to & \cA_-' =   - \bar v \cA_+ + \bar u \cA_- + a_-
\eea
where $a_\pm$ are complex constants parametrizing a  shift in addition to a pure $SU(1,1)$ transformation.
On the differentials $\partial_w\cA_\pm$ this induces a pure $SU(1,1)$ transformation
\bea
\label{dAtrans}
\partial_w\cA_+ & \to & \partial_w\cA_+'=  + u \partial_w\cA_+ - v \partial_w\cA_-
\no \\
\partial_w\cA_- & \to & \partial_w\cA_-' =   - \bar v \partial_w\cA_+ + \bar u \partial_w\cA_-
\eea
and implies that $\kappa^2$ and its complex conjugate are invariant under $SU(1,1)$.
Since $\cB$ is defined only up to a constant by (\ref{eq:comp}), the transformation of $\cA_\pm$ determines the  transformation of $\cB$ only up to a further constant shift. As discussed in \cite{D'Hoker:2016rdq}, however, for the transformation of the locally holomorphic data to induce the correct $SU(1,1)$ transformations on the supergravity fields, this shift has to vanish and we in addition have to require
\bea\label{eq:apm-cond}
a_- = \bar a_+
\eea
This condition is itself $SU(1,1)$ invariant and it implies that $\cG$ is invariant under (\ref{Atrans}) as well.
As a result, the metric functions $f_6^2$, $f_2^2$, $\rho^2$ are invariant, as expected for the metric in Einstein frame,
and the axion-dilaton scalar $B$ and gauge field $\cC$ transform as
\begin{align}
 B &\rightarrow B'=\frac{uB+v}{\bar v B+\bar u}
 \label{Bsu11}
 \\
 \cC &\rightarrow \cC ' = u \cC +v\bar \cC +\cC_0
 \label{Csu11}
\end{align}
Note that the transformation of $\cC$ includes a shift by a constant $\cC_0$ which can be compensated by a gauge transformation.

\subsubsection{Mapping to \texorpdfstring{$SL(2,\RR)$}{SL(2,R)}}

To translate the $SU(1,1)$ transformation of $B$ to the corresponding $SL(2,\mathds{R})$ transformation of $\tau$, we note that $B$ and $\tau$  are related by
\begin{align}
 B &= { \tau -i \over - \tau - i} = U(\tau) 
 &
 U &= { 1 \over \sqrt{-2i}} \begin{pmatrix} 1 & -i \\ -1 & -i \cr \end{pmatrix}
\end{align}
The normalization factor in $U$ has been chosen such that $\det U=1$.
The $SU(1,1)$ transformation in (\ref{Bsu11}) can be written as
\begin{align}\label{BV}
B &\to B' ={ u B + v \over \bar v B + \bar u} = V(B)
&
V &=  \begin{pmatrix}u & v \cr \bar v & \bar u\end{pmatrix}
\end{align}
with $|u|^2 -|v|^2=1$, while the $SL(2,\RR)$ transformation of $\tau$ is given by
\bea
\tau \to \tau ' = { a \tau + b \over c \tau + d} = M(\tau) \hskip 1in 
M =  \begin{pmatrix}a & b \\ c & d\end{pmatrix}
\eea
with $a,b,c,d \in \RR$ and $ad-bc=1$. The two transformations are related to one another by
\be
V(B) =B' =  U(\tau') = UM (\tau) = UMU^{-1} (B)
\ee
from which the identification of $SU(1,1)$ and $SL(2,\RR)$ parameters can be read off as 
\begin{align}
u & =  \half ( a+ib -i c +d)
&
v & =  \half (-a +ib +ic +d)
\label{uv-sl2r}
\end{align}
Given the relation between $B$ and $\p_w \cA_\pm$ in (\ref{2a6}) we find that the transformation of $B$ can be realized if the differentials $\p_w \cA_\pm$ are transformed as follows
\bea
 \begin{pmatrix}\p_w \cA_+  \\ \p_w \cA_-\end{pmatrix} \to  \begin{pmatrix}\p_w \cA_+'  \\ \p_w \cA_-'\end{pmatrix}
= 
(V^\dagger)^{-1} \begin{pmatrix}\p_w \cA_+  \\ \p_w \cA_-\end{pmatrix}
 \hskip 0.7in
(V^\dagger)^{-1} =  \begin{pmatrix}u & - v \\ - \bar v & \bar u\end{pmatrix}
\label{eq:dASU11}
 \eea

\subsection{Identification with 5-brane intersections}\label{sec:poles-5branes}

As discussed in detail in \cite{DHoker:2017mds}, the geometry of the supergravity solution close to a pole $p_\ell$ precisely matches the near-brane limit of the 5-brane solutions constructed in \cite{Lu:1998vh}.
In the notation of \cite{Lu:1998vh}, the charges of the 5-brane, $(q_1,q_2)Q$, are identified with the residue of $\partial_w\cA_+$ at the pole $p_m$, given by $Z_+^m$, via
\begin{align}
\label{eq:Zp-charge}
 (q_1-iq_2)Q&=\frac{8}{3}c_6^2Z_+^m
\end{align}
We note that in the convention of \cite{Lu:1998vh} $q_1$ corresponds to NS5 charge and $q_2$ to D5 charge,
and correspondingly $\Im(Z_+^m)$ translates to D5 charge while $\Re(Z_+^m)$ translates to NS5 charge.

\newpage

\section{Solutions with monodromy on the disc}
\setcounter{equation}{0}
\label{sec:sol}

In this section we will start from the global solutions without monodromy on the disc reviewed in the previous section and use them to construct physically regular solutions on a disc with punctures and non-trivial monodromy.
We will allow for an arbitrary number of punctures and for generic parabolic $SL(2,\RR)$ monodromies, as appropriate for the inclusion of 7-branes, but restrict the monodromies to be mutually commuting.
In sec.~\ref{sec:strategy} - \ref{sec:Gzero} we will detail the construction and derive the regularity conditions.
The results will be summarized in sec.~\ref{sec:summary}. In sec.~\ref{sec:param} and \ref{sec:d7} we will count the free parameters labeling distinct solutions and identify the punctures with \pqseven{} 7-branes.

\subsection{Strategy for solutions with monodromy}\label{sec:strategy}

Before discussing the construction of solutions with general monodromies in the upper half plane, we will outline the basic strategy for a simple example where we take $\Sigma$ to be a disc.
A general parabolic element of $SL(2,\mathds{R})$ can be parametrized by two real numbers $p,q$ as
\begin{align}\label{eq:Mpq}
 M_{[p,q]}=\begin{pmatrix} 1-p q & p^2\\ -q^2 & 1+p q\end{pmatrix}
\end{align}
and we will use this parametrization in the following.
The parameters of the corresponding $SU(1,1)$ transformation are given via (\ref{uv-sl2r}) by
\begin{align}
 u_{[p,q]}&=1+\frac{i}{2}(p^2+q^2)
 &
 v_{[p,q]}&=\frac{i}{2}(p-iq)^2
 \label{eq:uv-pq}
\end{align}
We will now consider the special case of a single puncture at the center of the disc, 
where only the axion has non-trivial monodromy and shifts by $1$.
The corresponding $SL(2,\mathds{R})$ matrix is
\be
\label{M}
M_{[1,0]} = \begin{pmatrix}1 & 1 \\ 0 & 1\end{pmatrix}
\ee
The entries of the corresponding $SU(1,1)$ transformation matrix $V$ in (\ref{BV}) are given by
$u=u_{[1,0]}$, $v=v_{[1,0]}$ with (\ref{eq:uv-pq}). 
The differentials correspondingly have to transform via (\ref{eq:dASU11}) as
\be
\p_w \cA_\pm & \to & \p_w \cA_\pm ' = \p_w \cA_\pm + { i \over 2} (\p_w \cA_+ - \p_w \cA_-)
\ee
To realize this monodromy around the point at the center of the disc, we introduce a coordinate such that $w=0$ corresponds to the center of the disc and $|w|=1$ to the boundary.
We may then realize the above monodromy by considering the logarithmic function, which has the appropriate monodromy as we wrap around the center by $w \to e^{2 \pi i } w$. 
Let $\p_w \cA_\pm^{(0)}$ be the differentials for a solution without monodromy on the disc, which are single-valued and meromorphic.
We then set 
\begin{align}
 \partial_w\cA_\pm&=\partial_w\cA_\pm^{(0)}+f_0\left(\partial_w\cA_+^{(0)}-\partial_w\cA_-^{(0)}\right)
 &
 f_0 (w)&=\frac{1}{4\pi}\ln w
\end{align}
The function $f_0$ is locally holomorphic on the disc, and this produces locally holomorphic differentials with the desired monodromy. 
What we have left to verify is that they satisfy the regularity conditions on $\kappa^2$ reviewed in sec.~\ref{sec:global-reg}. A straightforward calculation shows that
\begin{align}
\label{eq:kappa6}
 \kappa^2
 &=-|\partial_w\cA_+^{(0)}|^2+|\partial_w\cA_-^{(0)}|^2
 -(f_0+\bar f_0)\left|\partial_w\cA_+^{(0)}-\partial_w\cA_-^{(0)}\right|^2
\end{align}
The first term is positive in the interior of $\Sigma$ and vanishes on the boundary, since the differentials $\partial_w\cA_\pm^{(0)}$ were assumed to correspond to a regular solution.
For the second term we note that
\begin{align}
 -(f_0+\bar f_0)(w)&=-\frac{1}{4\pi}\ln|w|^2
\end{align}
is positive in the interior of the disc and vanishes on the boundary.
The second term in (\ref{eq:kappa6}) therefore is non-negative in the interior of the disc and zero on the boundary, such that $\kappa^2$ satisfies the regularity conditions in (\ref{2b1}) and (\ref{2b2}).

\subsection{The differentials \texorpdfstring{$\partial_w\cA_\pm$}{dA}}

We will now generalize the strategy outlined in the previous subsection to construct the differentials for an arbitrary number of punctures with commuting monodromies of the general form in (\ref{eq:Mpq}).
Instead of working with the disc, we will map to the upper half plane, so we can directly use the solutions of \cite{DHoker:2016ysh,DHoker:2017mds} reviewed in sec.~\ref{sec:review}.

\sm

The first step is to generalize the locally holomorphic function $f_0$ to the case with multiple punctures
with commuting monodromies at points $w_i$, $i=1,\cdots, I$ in the upper half plane. 
This is straightforward and yields
\begin{align}
f(w) &= \sum _{i=1}^I \frac{n_i^2}{4\pi} \ln \left ( \gamma_i\,{ w-w_i  \over w -\bar w_i} \right )
\label{f-def}
\end{align}
where $n_i\in\mathds{R}$ and $|\gamma_i|^2=1$ for $i=1,\cdots I$.
We note the following properties of the function~$f$
\begin{itemize}
 \item $f$ is locally holomorphic in the upper half plane, with branch points at $w_i$ around which
 \begin{align}
f(w_i + e^{2 \pi i} (w-w_i)) &= f(w) + {i \over 2} n_i^2
\end{align}
 \item the branch cuts associated with $w_i$ extend in a direction determined by $\gamma_i$ and can be parametrized as 
 \begin{align}\label{cuts}
w&=w_i+c\,\frac{\bar w_i- w_i}{c+\gamma_i} & c&\in [0,1]
\end{align}
in particular, $\gamma_i=+1$ and $\gamma_i=-1$ correspond  to a branch cut extending in the negative and positive imaginary direction, respectively;
 \item $-(f+\bar f)$ is positive in the interior of $\Sigma$ and vanishes on the boundary $\partial\Sigma$.
\end{itemize}
Using the function $f$ defined in (\ref{f-def}) and the differentials $\partial_w\cA_\pm^{(0)}$ for a solution without monodromy in the upper half plane, as given in (\ref{eq:A0}), we can now construct the differentials for a solution with axion monodromy in the upper half plane, by setting
\begin{align}
\label{eq:dA-ax}
 \partial_w\cA_\pm^\mathrm{ax}&=\partial_w\cA_\pm^{(0)}+f\left(\partial_w\cA_+^{(0)}-\partial_w\cA_-^{(0)}\right)
\end{align}
The monodromy of these differentials around $w_i$ is given by the $SU(1,1)$ transformation in (\ref{eq:dASU11}) with $u=u_{[n_i,0]}$, $v=v_{[n_i,0]}$ and (\ref{eq:uv-pq}).
This corresponds to the $SL(2,\RR)$ transformation
\begin{align}
 M_{[n_i,0]}&=\begin{pmatrix} 1 & n_i^2\\0 & 1\end{pmatrix}
\end{align}
thus realizing axion monodromies as desired. 

\sm

To generalize the construction to general parabolic $SL(2,\RR)$ monodromies of the form (\ref{eq:Mpq}),
we note that the transformation given in (\ref{eq:Mpq}) can be generated from $M_{[1,0]}$ given in (\ref{M})
by conjugating with an $SL(2,\RR)$ matrix $Q$ as follows
\begin{align}\label{eq:Q}
 M_{[p,q]}&=Q M_{[1,0]} Q^{-1}
 &
 Q&=\begin{pmatrix}p & -q/(p^2+q^2)\\[1mm]q & p/(p^2+q^2)\end{pmatrix}
\end{align}
To realize the transformation by $Q$ on the differentials in (\ref{eq:dA-ax}) we translate it to an $SU(1,1)$ transformation via (\ref{uv-sl2r}),
which yields
\begin{subequations}\label{dA-ansatz}
\begin{align}\label{eq:uqvq}
 u_Q&=\frac{1+\eta_+\eta_-}{2\eta_-}
 &
 v_Q&=\frac{1-\eta_+\eta_-}{2\eta_-}
 &
 \eta_\pm&=p\mp iq
\end{align}
Transforming the differentials (\ref{eq:dA-ax}) according to (\ref{dAtrans}) then yields differentials realizing
the desired monodromies, and with $\partial_w\cA_\pm \equiv (\partial_w\cA_\pm^\mathrm{ax})^\prime$ we find
\begin{align}\label{eq:dA-gen}
 \partial_w \cA_+&=+u_Q\partial_w\cA_+^{(0)}-v_Q\partial_w\cA_-^{(0)}+\eta_+f\left(\partial_w\cA_+^{(0)}-\partial_w\cA_-^{(0)}\right)
 \nonumber\\
 \partial_w \cA_-&=-\bar v_Q\partial_w\cA_+^{(0)}+\bar u_Q\partial_w\cA_-^{(0)}+\eta_-f\left(\partial_w\cA_+^{(0)}-\partial_w\cA_-^{(0)}\right)
\end{align}
\end{subequations}
This completes the construction of the differentials.
The expressions in (\ref{dA-ansatz}) realize $SL(2,\RR)$ monodromies 
\begin{align}\label{eq:Mpq-n}
 M_{[n_i p,n_i q]}&=\begin{pmatrix} 1-n_i^2 p q & n_i^2 p^2\\ -n_i^2q^2 & 1+n_i^2 p q\end{pmatrix}
\end{align}
around the points $w_i$ in the upper half plane, as desired. 
Moreover, since $\kappa^2$ is $SU(1,1)$ invariant 
and the differentials (\ref{dA-ansatz}) are obtained by an $SU(1,1)$ transformation from those in (\ref{eq:dA-ax}),
we have
\begin{align}\label{eq:kappa2}
 \kappa^2&=-|\partial_w\cA^\mathrm{ax}_+|^2+|\partial_w\cA^\mathrm{ax}_-|^2
 \no\\
 &=-|\partial_w\cA_+^{(0)}|^2+|\partial_w\cA_-^{(0)}|^2
 -(f+\bar f)\left|\partial_w\cA_+^{(0)}-\partial_w\cA_-^{(0)}\right|^2
\end{align}
Due to the properties of $f$ collected above, the differentials in (\ref{dA-ansatz}) therefore produce 
$\kappa^2$ that is positive in the interior of the upper half plane and zero on its boundary, thus satisfying the regularity conditions in (\ref{2b1}), (\ref{2b2}).
For any choice of global solution without monodromy, we therefore get suitable differentials for a solution with monodromy.

\sm

To facilitate the computations and arguments in the following sections, we will introduce a more convenient notation.
Namely, we split
\begin{align}\label{eq:dA-split}
\partial_w \cA_\pm&=\partial_w\cA_\pm^s + \eta_\pm\cF
\end{align}
where $\partial_w\cA_\pm^s$ denotes the single-valued part of the differentials
and the logarithmic part is denoted by $\cF$.
In terms of the seed solution without monodromy we have
\bea
\label{eq:dAsing}
 \partial_w\cA_+^s &=& +u_Q\partial_w\cA_+^{(0)}-v_Q\partial_w\cA_-^{(0)}
 \no \\
 \partial_w\cA_-^s &=& -\bar v_Q\partial_w\cA_+^{(0)}+\bar u_Q\partial_w\cA_-^{(0)}
\eea
for the single-valued part and the logarithmic part is given by
\begin{align}\label{eq:F}
 \cF&=f\left(\partial_w\cA_+^{(0)}-\partial_w\cA_-^{(0)}\right)
\end{align}
This can be spelled out more explicitly as
\begin{align}
\partial_w\cA_+^s (w) &=\sum_{\ell=1}^L\frac{Y_\pm^\ell}{w-p_\ell}
&
\cF(w) &=f(w) \sum_{\ell=1}^\ell \frac{Y^\ell}{w-p_\ell}
\end{align}
where we have defined convenient combinations of the residues as
\bea
\label{eq:Y}
Y_+^\ell &=& +u_Q Z_+^\ell-v_Q Z_-^\ell
\hskip 1in  
 Y^\ell=Z_+^\ell-Z_-^\ell
\no \\
Y_-^\ell&= & -\bar v_Q Z_+^\ell+\bar u_Q Z_-^\ell 
\eea
Due to the conjugation properties of $Z_\pm^\ell$ we have $\overline{Y_\pm^\ell}=-Y_\mp^\ell$ and that $Y^\ell$ is real.
Moreover, since the $Z_\pm^\ell$ sum to zero the same holds for $Y_\pm^\ell$ and we have $\sum_\ell Y_\pm^\ell=0$.

\subsection{The functions \texorpdfstring{$\cA_\pm$}{A}}\label{sec:cut}

\begin{figure}
\begin{center}
\begin{tikzpicture}[scale=0.9]
\shade [ top color=blue! 1, bottom color=blue! 30] (0.2,1.5)  rectangle (9.8,5);
\node at (9.5,4.5) {$\Sigma$};

\draw[dashed,very thick,black] (5.0,3) -- (5.0,5.0);
\draw[black] (5.0,3.0) node {$\bullet$};
\draw (5.0,2.7) node {$w_i$};

\draw[dashed,very thick,black] (7.5,3.8) -- (7.5,1.5);
\draw[black] (7.5,3.8) node {$\bullet$};
\draw (7.8,3.5) node {$w_j$};

\draw[black] (2.0,3.5) node {$\bullet$};
\draw (2.05,3.15) node {$w_k$};
\draw[dashed,very thick,black,domain=0:1,smooth,variable=\c] plot ({2-3.4641*\c/(\c*\c+\c+1)},{3.5-4*\c*(\c+0.5)/(\c*\c+\c+1)});

\draw [thick] (0,1.5) -- (10,1.5);
\draw (3,1.5) node{$\bullet$};
\draw (7,1.5) node{$\bullet$};
\draw (3,1.2) node{$p_{\ell'} $};
\draw (7,1.2) node{$p_\ell $};

\draw[ultra thick, red] plot [smooth] coordinates { (10,1.59) (7.9,1.8) (8.1,3.8) (7.5,4.3) (6.8,4.0) (5,2) (3,2.5)};
\draw [red] (3,2.5) node{$\bullet$};
\draw (3,2.2) node{$w $};
\end{tikzpicture}
\end{center}
\caption{Branch cuts for $\cF$ are drawn as black dashed lines and do not intersect each other or poles on the real line.
The cuts shown correspond to $\gamma_i=-1$, $\gamma_j=1$ and $\gamma_k=e^{i\pi/3}$.
An integration contour for $\cI$, which does not intersect any of the branch cuts, is shown in red.\label{fig:branchcut}}
\end{figure}
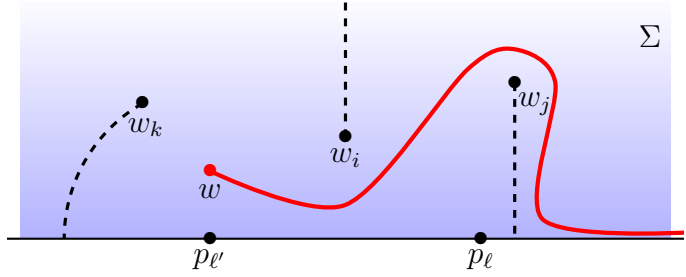

In this subsection we will construct the locally holomorphic functions $\cA_\pm$ from the differentials.
This in particular involves realizing a monodromy of the form (\ref{Atrans}) with the constant shifts related as in (\ref{eq:apm-cond}).
The differentials for solutions without monodromy could be integrated straightforwardly to obtain the locally holomorphic functions $\cA_\pm$.
For the differentials constructed in the previous section this is still possible, but due to the presence of the logarithms in $\cF$, their integrals involve dilogarithms. We find it more convenient to work with the integrals explicitly, and introduce the following notation
\begin{align}\label{eq:Asplit}
 \cA_\pm&=\cA_\pm^s + \eta_\pm \cI
\end{align}
We have once again separated the single-valued part $\cA_\pm^s$ from the part resulting from the logarithmic terms in the differentials $\eta_\pm\cI$. 
The expressions for $\cA_\pm^s$ are
\bea
 \cA_+^s&=& +u_Q\cA_+^{(0)}-v_Q\cA_-^{(0)}
 \no \\
 \cA_-^s&= & -\bar v_Q\cA_+^{(0)}+\bar u_Q\cA_-^{(0)}
\eea
with $\cA_\pm^{(0)}$ the holomorphic functions for the seed solution without monodromy as given in (\ref{eq:A0}).
More explicitly, we may write this as
\begin{align}
 \cA_\pm^s(w)&=\cA_\pm^0 +\sum_{\ell=1}^L Y_\pm^\ell \ln(w-p_\ell)
\end{align}
with the $Y_\pm^\ell$ given in (\ref{eq:Y}) and integration constants 
$\cA_\pm^0$, which are appropriate combinations of the integration constants of the seed solution without monodromy.
For the logarithmic part $\cI$ we have to discuss the choice of integration contour.
We will assume that all $\gamma_i$ and $w_i$ are chosen such that the resulting branch cuts do not intersect a pole on the real line,
and moreover that the branch cuts do not intersect each other in the interior of $\Sigma$.
The integration contour for $\cI$ with starting point at $+\infty+i0^+$ can then be chosen such that it does not intersect any of the branch cuts in~$\cF$. The contour is illustrated in fig.~\ref{fig:branchcut}.
The expression for $\cI$ becomes
\be\label{eq:I}
 \cI (w)=\int_\infty^w dz \cF(z)
\ee

\subsubsection{Behavior of \texorpdfstring{$\cA_\pm$}{A} across branch cuts}

\begin{figure}
\begin{center}
\begin{tikzpicture}[scale=0.9]
\shade [ top color=blue! 1, bottom color=blue! 30] (0.2,1.5)  rectangle (9.8,5);
\node at (9.5,4.5) {$\Sigma$};

\draw[dashed,very thick,black] (5.0,3) -- (5.0,5.0);
\draw[black] (5.0,3.0) node {$\bullet$};
\draw (5.0,2.6) node {$w_i$};

\draw[dashed,very thick,black] (7.5,3.8) -- (7.5,1.5);
\draw[black] (7.5,3.8) node {$\bullet$};
\draw (7.8,3.5) node {$w_j$};

\draw[black] (2.0,3.5) node {$\bullet$};
\draw (2.05,3.15) node {$w_k$};
\draw[dashed,very thick,black,domain=0:1,smooth,variable=\c] plot ({2-3.4641*\c/(\c*\c+\c+1)},{3.5-4*\c*(\c+0.5)/(\c*\c+\c+1)});

\draw [thick] (0,1.5) -- (10,1.5);
\draw (3,1.5) node{$\bullet$};
\draw (7,1.5) node{$\bullet$};
\draw (3,1.2) node{$p_{\ell'} $};
\draw (7,1.2) node{$p_\ell $};

\draw[ultra thick, red] (4.8,4.5) -- (4.8,3.0);
\draw[ultra thick, red] (5.2,4.5) -- (5.2,3.0);
\draw[ultra thick, red] (5.2,3.0) arc (-0:-180:0.2);

\draw[black, anchor=east] (5.08,4.5) node {$w-\epsilon$ $\bullet$};
\draw[black, anchor=west] (4.92,4.5) node {$\bullet$ $w+\epsilon$};

\node[red] at (5.6,3.7) {$C_2$};
\node[red] at (4.4,3.7) {$C_1$};

\end{tikzpicture}
\caption{Integration contour $C=C_1\cup C_2$, where $C_1$ denotes the left half of the contour shown in red and $C_2$ the right half.}
\label{fig:cont}
\end{center}
\end{figure}
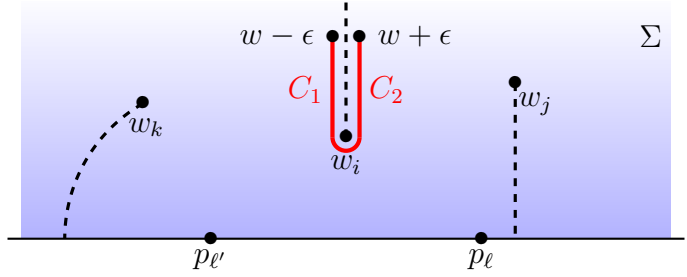

We can now evaluate the behavior of the holomorphic functions $\cA_\pm$ across the branch cut, for each cut individually.
Let $w$ be a point on the branch cut associated with a particular branch point $w_i$.
We can then evaluate the shift in the holomorphic functions by integrating around the branch cut as follows, 
\begin{align}
 \cA_\pm(w+\epsilon)-\cA_\pm(w-\epsilon)&=\int_{C}dz\, \partial_z\cA_\pm
 \nonumber\\
 &=\int_{C}dz\left[\partial_z\cA_\pm^s+\eta_\pm f\Big(\partial_z\cA_+^{(0)}-\partial_z\cA_-^{(0)}\Big)\right]
\end{align}
where the contour $C$ is illustrated in fig.~\ref{fig:cont} and the second equality follows using (\ref{eq:dA-split}), (\ref{eq:F}).
Since $\cA_\pm^s$ are holomorphic in the interior of $\Sigma$, the first term in square brackets cancels between
the segments $C_1$ and $C_2$ as $\epsilon\rightarrow 0$.
Moreover, again due to holomorphicity of $\partial_w\cA_\pm^{(0)}$, we can write the remaining part as
\be
 \cA_\pm(w+\epsilon)-\cA_\pm(w-\epsilon)=
 \int_{C_2}dz\,\eta_\pm (\Delta f) (\partial_z\cA_+^{(0)}-\partial_z\cA_-^{(0)})
\ee
where $\Delta f$ is the shift in $f$ across the branch cut.
Using $\Delta f=i n_i^2/2$, we then find
\be\label{eq:cut1}
 \cA_\pm(w+\epsilon)-\cA_\pm(w-\epsilon)=
\eta_\pm \frac{in_i^2}{2}\left[\cA_+^{(0)}(w)-\cA_-^{(0)}(w)-\cA_+^{(0)}(w_i)+\cA_-^{(0)}(w_i)\right]
\ee
The logarithmic singularity in the differentials $\partial_w\cA_\pm$ is integrable, and the functions $\cA_\pm$ therefore finite in the upper half plane. But they shift across the branch cut as given above.

The shift in eq.~(\ref{eq:cut1}) can be written as $SU(1,1)$ transformation supplemented by an additional complex shift as follows,
\begin{align}\label{eq:cut2}
 \cA_+(w+\epsilon)&= +u_{[n_i p,\,n_i q]} \cA_+(w-\epsilon)-v_{[n_i p,\,n_i q]}\cA_-(w-\epsilon)+a_+
 \nonumber\\
 \cA_-(w+\epsilon)&= -\bar u_{[n_i p,\,n_i q]} \cA_+(w-\epsilon)+\bar v_{[n_i p,\,n_i q]}\cA_-(w-\epsilon)+a_-
\end{align}
with the parameters $u$, $v$ as defined in (\ref{eq:uv-pq}). The complex constants $a_\pm$ are given by
\begin{align}\label{eq:apm}
 a_\pm&=-\eta_\pm \frac{in_i^2}{2}\left[\cA_+^{(0)}(w_i)-\cA_-^{(0)}(w_i)\right]
\end{align}
The $SU(1,1)$ parameters in (\ref{eq:cut2}) correspond to the $SL(2,\RR)$ transformation in (\ref{eq:Mpq-n}) and this is precisely the desired monodromy. 
To guarantee single-valued $\cG$ we only have to impose (\ref{eq:apm-cond}) on the shift parameter $a_\pm$.
Since we have $\eta_-=\bar\eta_+$, this condition amounts to the difference $\cA_+^{(0)}(w_i)-\cA_-^{(0)}(w_i)$ being imaginary
\begin{align}
 \cA_+^{(0)}(w_i)-\cA_-^{(0)}(w_i)+\mathrm{c.c.}&=0
 \label{eq:w1-0}
\end{align}
This can be expressed as a condition on the single-valued part of the differentials 
by noting that the residues are related by
\begin{align}\label{eq:etaY}
 \eta_- Y_+^k-\eta_+Y_-^k=Y^k
\end{align}
This yields the relation $\cA_+^{(0)}-\cA_-^{(0)}=\eta_-\cA_+^s-\eta_+\cA_-^s$ for the locally holomorphic functions
and we can express the condition in (\ref{eq:w1-0}) as
\begin{align}
 \eta_-\cA_+^s(w_i)-\eta_+\cA_-^s(w_i)+\mathrm{c.c.}&=0
 \label{eq:w1}
\end{align}
We have thus constructed the holomorphic functions $\cA_\pm$ for a solution with monodromy, 
and find that the location of the branch points is constrained by (\ref{eq:w1}).

\subsection{Regularity conditions for \texorpdfstring{$\cG$}{G}}\label{sec:Gzero}

We have constructed the differentials and the locally holomorphic functions $\cA_\pm$ and implemented the regularity conditions on $\kappa^2$.
It remains to implement the regularity conditions on $\cG$, which we will do in this section.
The positivity condition in the interior of $\Sigma$ in (\ref{2b1}) is automatically satisfied if we implement the condition $\cG=0$ on the boundary in (\ref{2b2}), for the same reasons as discussed in sec.~2.3 of \cite{DHoker:2017mds}.
Implementing $\cG=0$ on $\partial\Sigma$ proceeds in two steps. The first is to ensure that $\cG$ is piecewise constant along each boundary segment free of poles. The second is to then ensure that $\cG$ is also constant across poles.
The remaining free integration constant in $\cG$ (recalling its definition in terms of $\cB$ in (\ref{eq:comp}) and that $\cB$ is fixed only up to a constant) can then be used to set it to zero.

\subsubsection{Piecewise constant \texorpdfstring{$\cG$ on $\partial\Sigma$}{G on dSigma}}
\label{sec:Gconst}

Piecewise constant $\cG$ on the boundary $\partial\Sigma=\RR$ can be implemented by realizing a reflection symmetry across $\partial\Sigma$ on the locally holomorphic functions $\cA_\pm$ due to the fact that
\bea
\p_w \cG + \p_{\bar w} \cG & = & \Big ( \cA_+ (w) - \overline{\cA_- (w)} \Big )
\Big (  \p_w \cA_- (w) - \p_{\bar w} \cA_-   (\bar w) \Big ) 
\no \\ && 
- \Big ( \cA_- (w) - \overline{\cA_+ (w)} \Big )
\Big (  \p_w \cA_+ (w) - \p_{\bar w} \cA_+   (\bar w) \Big ) 
\eea
It is therefore sufficient to establish the conjugation property
\begin{align}\label{eq:Acc-cond}
\overline{ \cA_\pm (\bar w)} = - \cA_\mp (w)
\end{align}
which guarantees that $\partial_w\cG+\partial_{\bar w}\cG=0$ on $\partial\Sigma$ and hence that $\cG$ is piecewise constant.

\sm

To implement this conjugation property we start with the weaker condition on the derivatives.
The solution without monodromy is assumed to obey $\overline{\partial_{\bar w}\cA_\pm^{(0)}(\bar w)}=-\partial_w\cA_\mp^{(0)}(w)$, as discussed in \cite{DHoker:2017mds},
and from the explicit expressions for $\partial_w\cA_\pm^s$ in (\ref{eq:dAsing}) we see that the same is true for the single-valued part of the differentials.
From (\ref{eq:dA-split}) we therefore have
\begin{align}\label{eqn:dAcc1}
 \overline{\partial_{\bar w}\cA_\pm(\bar w)}&= \overline{\partial_{\bar w}\cA_\pm^s(\bar w)}+\overline{\eta_\pm\cF(\bar w)}
 \nonumber\\
 &=-\partial_w\cA_\mp^s(w)+\eta_\mp\overline{f(\bar w)} \left(-\partial_w\cA_-^{(0)}(w)+\partial_w\cA_+^{(0)}(w)\right)
\end{align}
To realize differentials with the desired conjugation property we therefore have to impose
\begin{align}\label{eq:fcc}
 \overline{f(\bar w)}  = - f(w)
\end{align}
With the symmetric distribution of the points $w_i$ and $\bar w_i$ under complex conjugation and the fact that the $\gamma_i$ are pure phases,
we indeed find from the definition of $f$ in (\ref{f-def}) that
\be
\overline{f(\bar w)} = \sum _{i=1}^I \frac{n_i^2}{4\pi} \, \overline{\ln \left ( \gamma_i{ \bar w-w_i  \over \bar w -\bar w_i} \right )}
= \sum _{i=1}^I \frac{n_i^2}{4\pi} \, \ln \left ( \frac{1}{\gamma_i}\,{ w- \bar w_i  \over w -w_i} \right )
=-f(w)
\ee
if the branch cut of the logarithm $\ln$ is chosen symmetrically with respect to complex conjugation.
With (\ref{eqn:dAcc1}) this yields the desired conjugation condition for the differentials
\begin{align}
 \overline{\partial_{\bar w}\cA_\pm(\bar w)}&=-\partial_w\cA_\mp(w)
\end{align}
Lifting this relation to the holomorphic functions $\cA_\pm$ now simply amounts to choosing the integration constants $\cA_\pm^0$ such that 
\begin{align}\label{eq:A0conj}
 \overline{\cA_\pm^0}&=-\cA_\mp^0
\end{align}
With the symmetric choice of branch cuts and the contour for $\cI_\pm$ in (\ref{eq:Asplit}), this suffices to ensure the conjugation property for the locally holomorphic functions $\cA_\pm$ in (\ref{eq:Acc-cond}),
and thus constant $\cG$ along each boundary component free of poles or branch cuts
($\cG$ also does not shift across branch cuts if the conditions (\ref{eq:w1}) are satisfied).

\subsubsection{Vanishing \texorpdfstring{$\cG$ on $\partial\Sigma$}{G on dSigma}}\label{sec:DeltaG}

Implementing the vanishing of $\cG$ amounts to realizing vanishing monodromy of $\cG$ around each pole.
Since we assumed that the branch cuts do not intersect the poles on the real axis, they will play a role only at the very end.
For the evaluation of the monodromy of $\cG$ around the pole $p_k$, $\Delta_k \cG$, we note that,
with a small $\epsilon\in \mathds{R}^+$,
\begin{align}
\Delta_k \cG & = 
 |\cA_+ (p_k- \ep)|^2 - |\cA_+ (p_k+ \ep)|^2
 - |\cA_- (p_k- \ep)|^2 + |\cA_- (p_k+ \ep)|^2
 \no \\ &
 +\Delta_k\cB+\Delta_k\bar\cB
 \label{eqn:DeltaG}
\end{align}
With $C_k$ a half circle  contour of radius $\epsilon$ centered on $p_k$ with counter-clockwise orientation,
the shift in $\cB$ is given by
\begin{align}
\Delta_k \cB &= \int _{C_k} dz\, \Big ( \cA_+  \, \p_z \cA_-  - \cA_-  \, \p_z \cA_+  \Big )
\label{eqn:DeltaB}
\end{align}
To evaluate the first line in (\ref{eqn:DeltaG}) explicitly, we note that we can evaluate the shift in $\cA_\pm$ across the pole by integrating the differentials along $C_k$, which yields
\begin{align}
\label{eq:cA-shift}
 \cA_\pm(p_k-\eps)-\cA_\pm(p_k+\eps)&=\int_{C_k}dw\,\partial_w\cA_\pm=i\pi\left[Y_\pm^k+\eta_\pm f(p_k)Y^k\right]
\end{align}
This also directly gives the residues of the differentials in the new solution at the poles on the real line.
Using that $\bar\eta_\pm=\eta_\mp$ and that $f$ is imaginary on $\partial\Sigma$, we find
\begin{align}\label{eq:deltaG-0}
 \Delta_k \cG &= i\pi\left[Y_+^k+\eta_+ f(p_k)Y^k\right]\left(\overline{\cA_+(p_k+\epsilon)}-\cA_-(p_k+\eps)\right)
 +\Delta_k\cB+\mathrm{c.c.}
\end{align}
Explicitly evaluating $\Delta_k\cB$ with its conjugate shows that it precisely reproduces the first term. 
We give the details of this calculation in app.~\ref{app:G0}.
Evaluating the first term in (\ref{eq:deltaG-0}) explicitly and using that $f(p_k)$ is imaginary, 
the resulting expression for the shift reads
\begin{align}
 \frac{\Delta_k\cG}{2\pi i}&=
 2\bar \cA^0 Y_+^k+2\cA^0 Y_-^k+\sum_{\ell\neq k} Y^{[\ell, k]}\ln |p_\ell-p_k|^2
 \nonumber\\&\hphantom{=}
 +Y^k\left(f(p_k)\Big[\eta_-\cA_+^s(p_k+\ep)-\eta_+\cA_-^s(p_k+\ep)\Big]-\cI(p_k+\ep)-\mathrm{c.c.}\right)
 \label{eq:DeltaG0b}
\end{align}
where $2\cA^0=\cA_+^0-\bar \cA_-^0$ and $Y^{[\ell, k]}=Y_+^\ell Y_-^k-Y_+^kY_-^\ell$.
The individual terms in the second line are divergent as $\epsilon\rightarrow 0$, but their combination is finite.
To make this manifest, it is convenient to perform an integration by parts in the expression for $\cI$. We relegate the details again to app.~\ref{app:G0}.
In the resulting expression we can then take $\epsilon\rightarrow 0$, as desired, and find
\begin{align}
 \frac{\Delta_k\cG}{2\pi i}&=
 2\bar \cA^0 Y_+^k+2\cA^0 Y_-^k+\sum_{\ell\neq k} Y^{[\ell, k]}\ln |p_\ell-p_k|^2
 \nonumber\\&\hphantom{=}
 +2f(p_k)Y^k(\eta_-\cA_+^0-\eta_+\cA_-^0)
 +Y^k\left(\int_\infty^{p_k}dw \sum_{\ell=1}^L Y^\ell \ln(w-p_\ell) \partial_w f-\mathrm{c.c.}\right)
 \label{eq:DeltaG0c}
\end{align}
As discussed in sec.~\ref{sec:cut} the integration contour has to be chosen in such a way that it does not cross any of the branch cuts,
and this is a natural form of the regularity conditions.
We can simplify the choice of contour by noting that $\partial_w f$ is meromorphic with simple poles in the upper half plane at the $w_i$,
such that the integrand in the second line is holomorphic except for at the poles of $\partial_w f$.
We can therefore also move the contour to the real line.
When deforming the integration contour shown in fig.~\ref{fig:branchcut} to approach the real line, 
we will only pick up the residues for the poles in $\partial_w f$ at those $w_i$ that are crossed.
This yields
\begin{align}
\int_\infty^{p_k}dw \ln(w-p_\ell)\partial_w f-\mathrm{c.c.}
&=
\int_\infty^{p_k} dx  \ln |x-p_\ell|^2 f^\prime(x)
+\sum_{i\in\cS_k} \frac{i n_i^2}{2} \ln |w_i-p_\ell|^2
\end{align}
where $\cS_k\subset\lbrace 1,\cdots, I\rbrace$ is the set of branch points for which the associated branch cut intersects the real line in the interval $(p_k,\infty)$ and the integral over $x$ is along the real line.
The explicit expression for $f'(x)$ reads
\be
f^\prime(x)=\sum_{i=1}^I\frac{i n_i^2}{2\pi}\frac{\Im(w_i)}{|x-w_i|^2}
\ee
The final form for $\Delta_k\cG$ can then be written as
\begin{align}
 \frac{\Delta_k\cG}{2\pi i}&=
 2\bar \cA^0 Y_+^k+2\cA^0 Y_-^k+\sum_{\ell\neq k} Y^{[\ell, k]}\ln |p_\ell-p_k|^2
 +2f(p_k)Y^k(\eta_-\cA_+^0-\eta_+\cA_-^0)
 \nonumber\\&\hphantom{=}
 +Y^k\sum_{\ell=1}^L Y^\ell\Bigg[\int_\infty^{p_k} dx \, f^\prime(x)  \ln |x-p_\ell|^2
 +\sum_{i\in\cS_k} \frac{i n_i^2}{2} \ln |w_i-p_\ell|^2\Bigg]
 \label{eq:deltaG}
\end{align}
To ensure that $\cG=0$ on the entire boundary of $\Sigma$ we have to enforce $\Delta_k\cG=0$ for all  $k=1,\cdots,L$.

\subsection{Summary of solutions and regularity conditions}\label{sec:summary}

We will now give a self-contained summary of the construction of solutions with monodromy
and of the regularity conditions, and discuss some additional points.
The data feeding into the construction are $L\geq 3$ poles $p_\ell$ on the real line,
$L-2$ zeros $s_n$ in the upper half plane and an overall complex normalization $\sigma$.
From those one constructs $Z_\pm^\ell$ via
\begin{align}
 Z_+^\ell  &=
 \sigma\prod_{n=1}^{L-2}(p_\ell-s_n)\prod_{k \neq\ell}^L\frac{1}{p_\ell-p_k}
 &
 Z_-^\ell&=-\overline{Z_+^\ell}
\end{align}
The additional data for the monodromies is given by a pair of real numbers $p,q$ and 
$I$ punctures $w_i$ in the upper half plane, with a real number $n_i$ 
for each puncture and a complex phase $\gamma_i$ fixing the direction of the branch cut.
This data fixes a function
\begin{align}
f(w) = \sum _{i=1}^I \frac{n_i^2}{4\pi} \ln \left ( \gamma_i\,{ w-w_i  \over w -\bar w_i} \right )
\end{align}
which encodes the branch points and additional branch cut structure.
Moreover, with
\begin{align}\label{eq:uqvq-rep}
 u_Q&=\frac{1+\eta_+\eta_-}{2\eta_-}
 &
 v_Q&=\frac{1-\eta_+\eta_-}{2\eta_-}
 &
 \eta_\pm&=p\mp iq
\end{align}
we define convenient shorthands for linear combinations of the $Z_\pm^\ell$ as
\begin{align}
Y_+^\ell &=+u_Q Z_+^\ell-v_Q Z_-^\ell
&
Y_-^\ell&=-\overline{Y_+^\ell}
 &
 Y^\ell&=Z_+^\ell-Z_-^\ell
\end{align}
The locally holomorphic functions for a solution with monodromy are then given by
\begin{align}\label{eqn:cA-summary}
 \cA_\pm&= \cA_\pm^0+\sum_{\ell=1}^L Y_\pm^\ell \ln(w-p_\ell) + \eta_\pm \int_\infty^w dz \;f(z)\sum_{\ell=1}^L \frac{Y^\ell}{z-p_\ell}
\end{align}
where $\cA_\pm^0$ are integration constants that are constrained by $\bar\cA_\pm^0=-\cA_\mp^0$.
The contour for the integral is chosen inside the upper half plane in such a way that it does not cross any of the branch cuts in $f$, as illustrated in fig.~\ref{fig:branchcut}.

\sm

The supergravity fields for these solutions have $SL(2,\RR)$ monodromies given by (\ref{eq:Mpq-n}) around the points $w_i$, as desired.
The residues of the differentials at the poles on the real line played a crucial role in the solutions without monodromy, for the identification with external 5-branes.
The differentials corresponding to $\cA_\pm$ in (\ref{eqn:cA-summary}) are given by
\begin{align}\label{eqn:dcA-summary}
 \partial_w\cA_\pm&= \sum_{\ell=1}^L \frac{Y_\pm^\ell+\eta_\pm f(w) Y^\ell}{w-p_\ell}
\end{align}
where we note that the numerators are non-trivial functions of $w$.
The residues of these differentials at the poles on the real line appeared already in (\ref{eq:cA-shift}), and are given by
\be\label{eq:cY}
\cY_\pm^\ell=Y_\pm^\ell+\eta_\pm f(p_\ell)Y^\ell
\ee
It is these residues that correspond to the charges of the external 5-branes via the identification reviewed in sec.~\ref{sec:poles-5branes}: since the match of the geometry close to a pole to a 5-brane solution only uses the local form
of the solution around the pole, this match carries over to the solution with monodromy straightforwardly.
We will therefore use the $\cY_\pm^\ell$ as shorthand for the combination in (\ref{eq:Y}) whenever convenient.

\sm

The parameters introduced above are constrained by regularity requirements and therefore not all independent.
The construction already ensures that the regularity conditions on $\kappa^2$ are satisfied, but to have single-valued $\cG$ which vanishes on the boundary the parameters in addition have to be chosen such that eqs.~(\ref{eq:w1}) and (\ref{eq:deltaG}) are satisfied.
Using the conjugation condition $\bar\cA_\pm^0=-\cA_\mp^0$ and the relation (\ref{eq:etaY}), 
we can write these conditions more explicitly as
\begin{align}
\label{eq:w1-summary}
 0&=2\eta_-\cA_+^0-2\eta_+\cA_-^0+\sum_{\ell=1}^LY^\ell \ln|w_i-p_\ell|^2
 &i&=1,\cdots,I
 \\
 0&=2\cA_+^0\cY_-^k-2\cA_-^0\cY_+^k
 +\sum_{\ell\neq k} Y^{[\ell, k]}\ln |p_\ell-p_k|^2+Y^kJ_k
 &
 k&=1,\cdots,L
 \label{eq:DeltaG0-summary}
\end{align}
where $Y^{[\ell, k]}=Y_+^\ell Y_-^k-Y_+^kY_-^\ell$.
With $\cS_k\subset\lbrace 1,\cdots, I\rbrace$ denoting the set of branch points for which the associated branch cut intersects the real line in the interval $(p_k,\infty)$, $J_k$ is given by
\begin{align}
 J_k&=\sum_{\ell=1}^L Y^\ell\Bigg[\int_\infty^{p_k} dx f^\prime(x)  \ln |x-p_\ell|^2
 +\sum_{i\in\cS_k} \frac{i n_i^2}{2} \ln |w_i-p_\ell|^2\Bigg]
\end{align}
 where the integral over $x$ is along the real line.
The conditions in (\ref{eq:DeltaG0-summary}) ensure that the shift in $\cG$ across the pole $p_k$, $\Delta_k\cG$, vanishes, while those in 
(\ref{eq:w1-summary}) ensure that $\cG$ is continuous across the branch cuts associated with $w_i$.
In contrast to the case without monodromy, the sum over the regularity conditions in (\ref{eq:DeltaG0-summary}) does not manifestly vanish, and we therefore in general have $L$ independent conditions.
However, satisfying the branch point conditions in (\ref{eq:w1-summary}) does imply that $\sum_k \Delta\cG=0$, and consequently that the sum over the conditions in (\ref{eq:DeltaG0-summary}) vanishes:
By the arguments of sec.~\ref{sec:Gconst}, $\cG$ is constant along each boundary segment free of poles.
Therefore, since $\sum_k \Delta\cG$ gives the total change in $\cG$ across all poles, it must equal the shift 
in $\cG$ across all branch cuts.
Satisfying (\ref{eq:w1-summary}) for each branch point implies that $\cG$ is continuous across all branch cuts and  therefore $\sum_k \Delta\cG=0$.

\subsection{Counting free parameters}\label{sec:param}

Having gathered the parameters and the constraints on the parameters for general solutions to be regular with monodromy in a convenient form, we can now count the moduli.
The parameters associated with the $Z_\pm^\ell$ are
\bea
s_n && 2L-4\hbox{ real parameters}\no\\
\sigma \ && 2\hbox{ real parameters}\no\\
p_\ell && L\hbox{ real parameters}
\label{eq:paramcount1}
\eea
The remaining parameters are the integration constants $\cA_\pm^0$, which are related by the conjugation condition (\ref{eq:A0conj}) and therefore correspond to only 2 real parameters, and the parameters directly associated with the punctures and monodromies. 
Namely,
\bea
\cA^0_\pm&&2\hbox{ real parameters}\no\\
p,q && 2\hbox{ real parameters}\no\\
\omega_i && 2I \hbox{ real parameters}\no\\
n_i && I \hbox{ real parameters}\no\\
\gamma_i && I \hbox { real parameters}
\label{eq:paramcount2}
\eea
Altogether, (\ref{eq:paramcount1}) and (\ref{eq:paramcount2}) are $3L+4I+2$ real degrees of freedom.
Those have to satisfy the $L+I-1$ independent conditions in (\ref{eq:w1-summary}) and (\ref{eq:DeltaG0-summary}).
We also have to account for the redundancy due to the $SL(2,\mathds{R})$ automorphisms of the upper half plane, which map to equivalent solutions and can be used to fix, e.g., the position of three of the poles at will.
Moreover, one of the parameters in $(p,q,n_i)$ is redundant, since an overall rescaling of $p$ and $q$ can be compensated by rescaling the $n_i$.
We are thus left with 
\begin{align}
 2L-1+3I
\end{align}
free real parameters.
As discussed in \cite{DHoker:2017mds}, the general $L$-pole solution without monodromy has $2L-2$ free real parameters.
Compared to the solution with no monodromy, each branch point therefore adds three real degrees of freedom,
and we in addition have one extra free parameter.
The extra parameter corresponds to the choice of $SL(2,\RR)$ monodromy that is fixed by $(p,q,n_i)$.
With the $n_i$ unconstrained we can take it e.g.\ as the the phase of $p-iq$.
For $I=0$ the dependence on that extra parameter becomes trivial, and the parameter count therefore reduces to the expected number for a solution without monodromy.

\subsection{Identification of punctures with \texorpdfstring{\pqseven{}}{[p,q]} 7-branes}\label{sec:d7}

In this section we will discuss the identification of the punctures $w_i$ with the location of \pqseven{} 7-branes.
The monodromies around the punctures in (\ref{eq:Mpq-n}) are precisely those expected for a \pqseven{} 7-brane \cite{Gaberdiel:1997ud}, which certainly suggests this identification. We will discuss this in more detail by explicitly working out the form of all supergravity fields near the $w_i$.
It will be sufficient to fix \pqseven$=[1,0]$ and discuss the relation of the punctures in the resulting solution to D7-branes. Since the solutions with general \pqseven{} monodromies were obtained from those with $[1,0]$ monodromies by an $SL(2,\RR)$ transformation, and the \pqseven{} 7-branes are related to $[1,0]$ 7-branes by the same $SL(2,\RR)$ transformation,  the identification directly extends to general \pqseven{} once it is established for $[1,0]$.
In sec.~\ref{sec:asympt} we will analyze the asymptotic behavior of the supergravity fields near the $w_i$ for \pqseven$=[1,0]$ and in sec.~\ref{sec:match} we will compare to the expected behavior for D7-branes.

\subsubsection{Asymptotic behavior near \texorpdfstring{$[1,0]$}{[1,0]} branch points}\label{sec:asympt}

For $\pqseven{}=[1,0]$ we have $u_Q=1$, $v_Q=0$, $\eta_\pm=1$, and the expressions simplify correspondingly.
In particular, $\partial_w\cA_\pm^s=\partial_w\cA_\pm^{(0)}$.
To analyze the metric factors near $w_i$ we need the behavior of $\kappa^2$, $\cG$ and $\partial_w\cG$.
We introduce a coordinate $\xi$ centered on the branch point
\be
\xi=\gamma_i\frac{w-w_i}{w-\bar w_i}
\ee
and will assume $|\xi|\ll 1$ for the expansions.
For the asymptotic behavior of $\kappa^2$ we then find
\begin{align}\label{kappa2a}
\kappa^2&=-\frac{n_i^2}{4\pi}|c|^2\ln |\xi|^2 +\mathcal O(1)
&
c&=\partial_w\cA_-^{(0)}-\partial_w\cA_+^{(0)}\Big\vert_{w=w_i}
\end{align}
For the expansion of $\cG$ it is convenient to separate off the contribution purely from the $\cA_\pm^{(0)}$
as $\partial_w\cG^{(0)}$, so we have
\be
\partial_w\cG
=\partial_w\cG^{(0)} + c(\cI-\bar \cI) + \cF(\bar\cA_+^{(0)}-\cA_-^{(0)}+\cA_+^{(0)}-\bar \cA_-^{(0)})
\ee
Due to (\ref{eq:w1}), the term multiplying $\cF$ is $\mathcal O(\xi)$. 
Moreover, due to the same condition
$\partial_w\cG^{(0)}=c(\cA_+^{(0)}-\bar\cA_-^{(0)})+\mathcal O(\xi)$.
Therefore,
\begin{align}
\label{eq:dGexp}
\partial_w\cG&=g_1 + \mathcal O(\xi\ln|\xi|^2)
&
g_1&=c(\cA_+-\bar\cA_-)\Big\vert_{w=w_i}
\end{align}
Note that $\cI$ and thus $\cA_\pm$ are finite at $w_i$, so $\partial_w\cG$ is finite as $w\rightarrow w_i$ as well.
Upon integrating the same applies for $\cG$, and to conveniently collect the $\mathcal O(1)$ terms we use
\be
\cG=g_0+\mathcal O(\xi)
\ee
Due to the regularity condition $\cG>0$ in $\Sigma$ we have $g_0\in\mathds{R}^+$, 
while $g_1$ is not constrained, $g_1\in\mathds{C}$.
Using the definition of $R$ yields
\be
R=\frac{4\pi |g_1|^2}{n_i^2 g_0 |c|^2(-\ln|\xi|^2)}+\mathcal O\left((\ln|\xi|)^{-2}\right)
\ee
and therefore $R\rightarrow 0^+$ as $w\rightarrow w_i$.
With the expression for the metric factors in (\ref{eq:f2f6rho}), we then find,
to leading order in the $\xi$-expansion,
\begin{align}
f_6^2&\approx c_6^2\sqrt{6g_0}
&
f_2^2&\approx\frac{c_6^2}{9}\sqrt{6g_0}
&
\rho^2&\approx\frac{n_i^2}{4\pi}\frac{|c|^2}{\sqrt{6g_0}}(-\ln|\xi|^2)
\end{align}
That is, in Einstein frame the radii of $AdS_6$ and $S^2$ are finite, while $\rho^2$ diverges logarithmically.
Note that the sign ensures that $\rho^2$ is positive.
With $|dw|^2\approx |w_i-\bar w_i|^2 |d\xi|^2$ and (\ref{2a1}), the complete metric takes the form
\begin{align}
ds^2 &\approx
c_6^2\sqrt{6g_0}\left( ds^2 _{AdS_6} + \frac{1}{9}\, ds^2 _{S^2}\right) 
+\frac{n_i^2 |c|^2}{\pi \sqrt{6g_0}}(-\ln\left|\xi\right|^2)\, |w_i-\bar w_i|^2 |d\xi|^2
\label{eqn:metric-w1}
\end{align}

\sm

To derive the expansion of $B$ it is convenient to rewrite (\ref{2a6}) as
\be
B=-1+\frac{(\partial_w\cA_+-\partial_w\cA_-)\partial_{\bar w}\cG+R(\partial_{\bar w}\bar\cA_+-\partial_{\bar w}\bar \cA_-)\partial_w\cG}{R\partial_{\bar w}\bar\cA_+\partial_w\cG-\partial_w\cA_-\partial_{\bar w}\cG}
\ee
Since $\partial_w\cA_+-\partial_w\cA_-=\partial_w\cA^{(0)}_+-\partial_w\cA^{(0)}_-$, the numerator in the second term is $\mathcal O(1)$, while the denominator is $\mathcal O(\ln|\xi|^2)$.
The explicit expansion reads
\be
 B=-1+\frac{c}{\cF}+\mathcal O\left((\ln|\xi|)^{-2}\right)
\ee
The expansion for $\tau$ is conveniently derived using $\tau=-i+2i/(1+B)$, which yields
\be
\tau=-\frac{in_i^2}{2\pi}\ln\xi + \tau_0
\label{eqn:tau-w1}
\ee
where $\tau_0$ is finite at $\xi=0$ and single-valued up to terms of $\mathcal O(1/\ln|\xi|)$.
With $\tau=\chi+ie^{-2\phi}$ we find the explicit expressions for axion and dilaton, to leading order near the branch point,
\begin{align}
\chi&\approx-\frac{i n_i^2}{4\pi}\left(\ln\xi-\overline{\ln\xi}\right)+\chi_0
&
e^{-2\phi}&\approx -\frac{n_i^2}{4\pi}\ln|\xi|^2
\end{align}
where $\chi_0$ is finite at $\xi=0$ and single-valued up to terms of $\mathcal O(1/\ln|\xi|)$.
We therefore find the expected axion monodromy $\chi\rightarrow \chi+n_i^2$ when encircling $w_i$ counterclockwise at an infinitesimal radius.
Moreover, we see that the exponentiated dilaton diverges logarithmically.

\sm

To derive the form of $\cC$ near the branch point we start from the expression in (\ref{2a7}).
With (\ref{eq:dGexp}) and (\ref{eq:w1}), one finds that 
$\partial_w \cG \, \p_{\bar w} \bar \cA_- +  \partial_{\bar w}  \cG \, \p_w \cA_+=\mathcal O(1)$.
The second term in the bracket of (\ref{2a7}) therefore is $\mathcal O((\ln|\xi|)^{-2})$.
Up to terms of $\mathcal O((\ln|\xi|)^{-2})$, the behavior of $\cC$ near the branch point is
thus given by
\bea
 \cC \approx \frac{4 i c_6^2}{9}\left (  
 - \bar  \cA_- - 2 \cA_+
 +
\frac{\overline{\ln\xi}}{\ln|\xi|^2}(\cA_+-\bar\cA_-)  \right )
\label{eq:Cexp}
\eea
The two-form potential is therefore finite at the branch point but not necessarily single-valued across the branch cut.
We note that, due to (\ref{eq:w1}), $\cA_+-\bar\cA_-$ is imaginary at $w_i$.
Since the monodromy of $\ln\xi$ is imaginary as well, we find that the real part of $\cC$ is 
single-valued and only the imaginary part is in general not.

In general, $C_{(2)}$ and correspondingly $\cC$ transform non-trivially under $SL(2,\RR)$, as given in (\ref{Csu11}).
For the monodromy considered here we would expect the imaginary part of $\cC$ to receive a shift proportional to the real part of $\cC$.
However, since $\ln\xi / \ln |\xi|^2\rightarrow 0$ as the branch point is approached, the expansion in (\ref{eq:Cexp}) shows that the shift in $\cC$ vanishes when encircling the branch point at an infinitesimal radius.
This reveals the constant gauge transformation in (\ref{Csu11}) as
\be
\cC_0=-\frac{in_i^2}{2}\big(\cC(w_i)+\bar \cC(w_i)\big)
\ee

\subsubsection{Matching to 7-branes}\label{sec:match}

With the asymptotic behavior of the solution with $[1,0]$ monodromy near the branch point in hand, we can now attempt a physical interpretation.
The form of the monodromy clearly suggests that the branch points correspond to D7-branes, 
and we will now extend the discussion to include all supergravity fields.
The D7-brane solution has been worked out already in \cite{Greene:1989ya}, 
but we will take it in the form given in \cite{Ortin:2015hya}.
To match to \cite{Ortin:2015hya}, we rewrite the metric near the branch point,
as given in (\ref{eqn:metric-w1}), as
\begin{align}
ds^2 &\approx
c_6^2\sqrt{6g_0}\left( ds^2 _{AdS_6} + \frac{1}{9}\, ds^2 _{S^2}\right) 
+\Im(\cH) |dz |^2 
&
\cH=-\frac{in_i^2}{2\pi}\ln z
\end{align}
where we changed coordinates to $z=c\,|w-w_i|\,\xi$. The axion-dilaton $\tau$ near the branch point, as given in (\ref{eqn:tau-w1}), then takes the form $\tau \approx\cH +\tilde\tau_0$,
where $\tilde\tau_0$ is finite at $z=0$ and single-valued up to terms of $\mathcal O(1/\ln|z|)$.

\sm

The metric of the transverse space parametrized by $z$ immediately matches the form of the flat-space D7-brane solution given in  (19.74), (19.75) of \cite{Ortin:2015hya}, taking into account the difference in conventions for the spacetime signature.
The (trivial) scaling of the remaining part of the metric with $z$ in Einstein frame also agrees with the flat-space D7-brane solution, but with  $AdS_6\times S^2$ replacing $\mathds{R}^{1,7}$.
The axion-dilaton $\tau$ matches up to the  finite offset $\tilde \tau_0$, and for $n_i=1$ we  find the same monodromy.
The two-form gauge field is generically non-vanishing at $w_i$, which is another difference to the flat-space D7-brane solution.
We therefore find a D7-brane in a non-trivial background, where the axion-dilaton and the two-form fields have non-trivial background values and the D7-brane wraps $AdS_6\times S^2$.

\sm

The stronger background dependence exhibited by a D7-brane compared to the virtual background independence observed near any of the semi-infinite 5-branes (as discussed  in \cite{DHoker:2017mds}) can be understood from the behavior of $\rho^2$. Close to the poles on the real line, where the semi-infinite 5-branes reside, the metric factor $\rho^2$ behaves as $\mathcal O(r^{-3/2}|\ln r|^{-1/4})$. Therefore, the metric distance of any interior point of $\Sigma$ to the location of the pole is infinite. 
This offers the possibility to move out on each of the semi-infinite 5-branes of  the web and decouple from the intersection. Close to the branch point, however, $\rho^2$ only diverges logarithmically, so the proper distance to other points in $\Sigma$ remains finite.
We can not move away from the intersection to a point where the 5-branes decouple.
We will expand on the interpretation in the context of 5-brane webs in sec.~\ref{sec:5-brane-webs}.

\sm

The asymptotic behavior of the supergravity fields for a branch point with generic parabolic \pqseven{} monodromy can be obtained by an $SU(1,1)$ transformation with parameters given in (\ref{eq:uqvq}) from the results in sec.~\ref{sec:asympt}.
The Einstein-frame metric is invariant while the axion-dilaton $B$ and the gauge field $\cC$ transform as in (\ref{Bsu11}), (\ref{Csu11}).
Since \pqseven{} 7-branes are obtained from D7-branes precisely by the $SL(2,\RR)$ transformation corresponding to this element of $SU(1,1)$, this straightforwardly extends the above discussion to generic $p,q$.

\newpage

\section{Example solutions with monodromy}
\setcounter{equation}{0}
\label{sec:examples}

In this section we will explicitly construct example solutions with monodromy and illustrate that the regularity conditions derived in the previous sections are indeed sufficient to guarantee smooth supergravity solutions with the desired monodromies.
We will also explicitly exhibit the real degree of freedom in choosing the position of the 7-branes.

\sm

The simplest case to consider are 3-pole solutions. 
3-pole solutions without 7-branes are all $SL(2,\mathds{R})$ dual to each other up to an overall rescaling of the charges, as discussed in \cite{DHoker:2017mds}.
This is to be expected already from the parameter count: For solutions without monodromy there are $4$ independent parameters after taking into account the redundancy due to the $SL(2,\mathds{R})$ automorphisms of the upper half plane. These parameters are further reduced by the $SL(2,\mathds{R})$ duality transformations of Type IIB supergravity to a single parameter corresponding  to the overall scale of the residues.
For solutions with monodromy, however, this is not true anymore. 
For solutions with $I\geq 1$ branch points, there are $5+3I$ free parameters according to the counting in sec.~\ref{sec:param} and $2+3I$ after taking into account the $SL(2,\mathds{R})$ duality transformations of Type IIB supergravity.
So the 3-pole solutions already yield families of inequivalent solutions and we will discuss some of the features in the following.

\subsection{3-pole solutions with D7 and D5-branes}\label{sec:3-pole-ex-2}

We will start with a simple example where the regularity conditions can be solved straightforwardly in closed form, to illustrate the procedure and discuss some general points.
We will consider the case where a solution with D7-brane monodromy is constructed from a 3-pole solution where one of the poles corresponds to D5-branes. Recalling the discussion in sec.~\ref{sec:poles-5branes} that means the corresponding residue $Z_+^\ell$ is purely imaginary.
By $SL(2,\RR)$ duality the discussion extends straightforwardly to the case where generic 7-brane charges coincide with the charges of one of the 5-branes, but to keep the expressions simple we fix them as corresponding to D7 and D5-branes. In that case we have $v_Q=0$ and $u_Q=\eta_\pm=1$.

\sm

For 3-pole solutions the $SL(2,\mathds{R})$ automorphisms of the upper half plane can be used to fix the location of all poles,
and we will use
\begin{align}
p_1&=1 & p_2&=0 & p_3&=-1
\label{eq:3-pole-pos}
\end{align}
We will take the pole $p_1$ to correspond to a stack of D5-branes.
Since the residues in the seed solutions sum to zero, this constrains the real parts of the other two residues to sum to zero,
and we have
\begin{align}\label{eq:D5D7-res}
Z_+^1&=i N
&
Y^1&=0
&
Y^2&=-Y^3
\end{align}
with $N\in\mathds{R}\setminus\lbrace 0\rbrace$.
This simplifies the regularity conditions (\ref{eq:w1-summary}), (\ref{eq:DeltaG0-summary})  considerably.
$2+I$ of these conditions are independent and have to be solved.
The condition in (\ref{eq:DeltaG0-summary}) for $k=1$ fixes the real part of the integration constants $\cA_\pm^0$ as
\be\label{eq:A0-D5}
\cA_+^0-\cA_-^0=-Y^3\ln 2
\ee
Recall that the integration constants are related by the conjugation condition (\ref{eq:A0conj}).
With that real part fixed we can solve the branch point conditions (\ref{eq:w1-summary}), which imply
\begin{align}\label{eq:D5D7-wi}
 w_i&=\frac{1}{3}\left(1+2e^{i\alpha_i}\right)
 &0\leq \alpha_i\leq \pi
\end{align}
The location of the branch points is thus constrained to a half circle of radius $2/3$ centered on the real line at $1/3$.
It intersects the real line at the location of the pole $p_1$, corresponding to D5-branes, and at $-1/3$, in between the other two poles.
The remaining regularity conditions are the conditions in (\ref{eq:DeltaG0-summary}) for $k=2,3$.
Since we solved the branch point conditions in (\ref{eq:w1-summary}), these remaining conditions are not linearly independent and solving one of them implies the other one. 
We therefore find only one more real constraint, fixing the imaginary part of $\cA_\pm^0$ which was left unconstrained by (\ref{eq:A0-D5}). This yields
\be
\cA_+^0=\frac{1}{2}J_2-\frac{Y^3\cY_+^2}{Y^2}\ln 2
\ee
The combination of this $\cA_+^0$ with $w_i$ in (\ref{eq:D5D7-wi}) solves all the regularity conditions (\ref{eq:w1-summary}), (\ref{eq:DeltaG0-summary}).

\sm

The regularity conditions do not fix $n_i$ and $\gamma_i$, and the curve on which the branch points can be placed is independent of both parameters. In addition we have one real parameter $\alpha_i$ for each puncture, specifying the position of the branch point on the curve in $\Sigma$. This clearly exhibits the $3$ extra parameters introduced by each branch point, in line with the discussion in sec.~\ref{sec:param}.
The additional parameters associated with the branch points do affect the residues $\cY_\pm^\ell$ of the differentials at the poles on the real line, as given in (\ref{eq:cY}).
With (\ref{eq:D5D7-res}) they explicitly read
\begin{align}
  \cY_+^1&=Z_+^1 & \cY_+^2&=Z_+^2+f(p_2)Y^2 & \cY_+^3&=Z_+^3-f(p_3)Y^2 
\end{align}
Since $f$ is imaginary on the real line, the residue at each pole changes by an imaginary amount proportional to the real part of the residue. 
That is, the D5 charge of the 5-brane changes by an amount proportional to its NS5 charge.
In particular, the residue at $p_1$, corresponding to the D5 charge there, is unaffected by the addition of the D7-branes.
The total charge non-conservation is given by $\sum_\ell \cY_+^\ell=(f(p_2)-f(p_3))Y^2$.
It is independent of the choice of $\gamma_i$, but varies with $n_i$ and $\alpha_i$.

\sm

Regarding the choice of orientation for each branch cut, one can realize ``topologically'' different configurations, by choosing different pairs of adjacent poles between which the branch cut intersects the real line. These different configurations have an immediate interpretation from the brane intersection picture,  namely as the choice of semi-infinite external branes between which the branch cut is located. The phases $\gamma_i$ fixing the orientation of the branch cuts, however, can be varied continuously. 
 Indeed, fixing all other parameters and varying one of the $\gamma_i$ such that the associated branch cut varies without crossing any of the poles, we find a linear dependence of the residues $\cY_+^\ell$ on $\arg(\gamma_i)$.
 The change in the residue $\cY_+^k$ as the branch cut associated with $w_i$ crosses 
 the pole $p_k$ is discrete and given by $\Delta\cY_+^k=\frac{i}{2}n_i^2Y^k$.
 We will come back to the interpretation of the continuous moduli in the brane web picture in sec.~\ref{sec:5-brane-webs}.
 As a last point, we note that the solution without monodromy can be recovered if the branch cuts are chosen e.g.\ along the negative imaginary direction and the branch points are moved along the allowed curve in (\ref{eq:D5D7-wi}) to approach the real line at $-1/3$.
 At the real line  the $w_i$ ``annihilate'' with their mirror points in the lower half plane, leading back to a solution without monodromy.

\subsection{3-pole solution with \texorpdfstring{$[1,0]$}{[1 0]} branch point}\label{sec:3pole-1cut}

To illustrate that the constructions outlined in sec.~\ref{sec:review} indeed yield solutions with the desired monodromies and regularity properties, we will now show explicit plots for a generic solution with three poles and one puncture corresponding to a D7-brane.
We fix the poles again as in (\ref{eq:3-pole-pos}).
As an explicit example we start from the 3-pole solution discussed in sec.~4.1 of \cite{DHoker:2017mds}, for which the zero in the upper half plane and $\sigma$ were chosen as
\begin{align}\label{eq:3pole-ex}
s&=\frac{1}{2}+2i
&
\sigma&=i
\end{align}
Plots of the solution without punctures were shown in sec.~4 of \cite{DHoker:2017mds}.
Adding 7-branes introduces additional parameters $(w_i, n_i,\gamma_i)$ as well as the charges $p$, $q$. 
We add a single D7-brane with $\pqseven{}=[1,0]$, such that $u_Q=\eta_\pm=1$, $v_Q=0$, and fix
\begin{align}
n_1&=1 & \gamma_1&=-1
\end{align}
The regularity conditions in (\ref{eq:DeltaG0-summary}) for $k=1,2$ can be solved for $\cA_\pm^0$ straightforwardly, and as the remaining independent constraint we can then take the condition associated with the branch point in (\ref{eq:w1-summary}).
That constrains the location of the D7-brane.
\begin{figure}
  \centering
  \includegraphics[height=42mm]{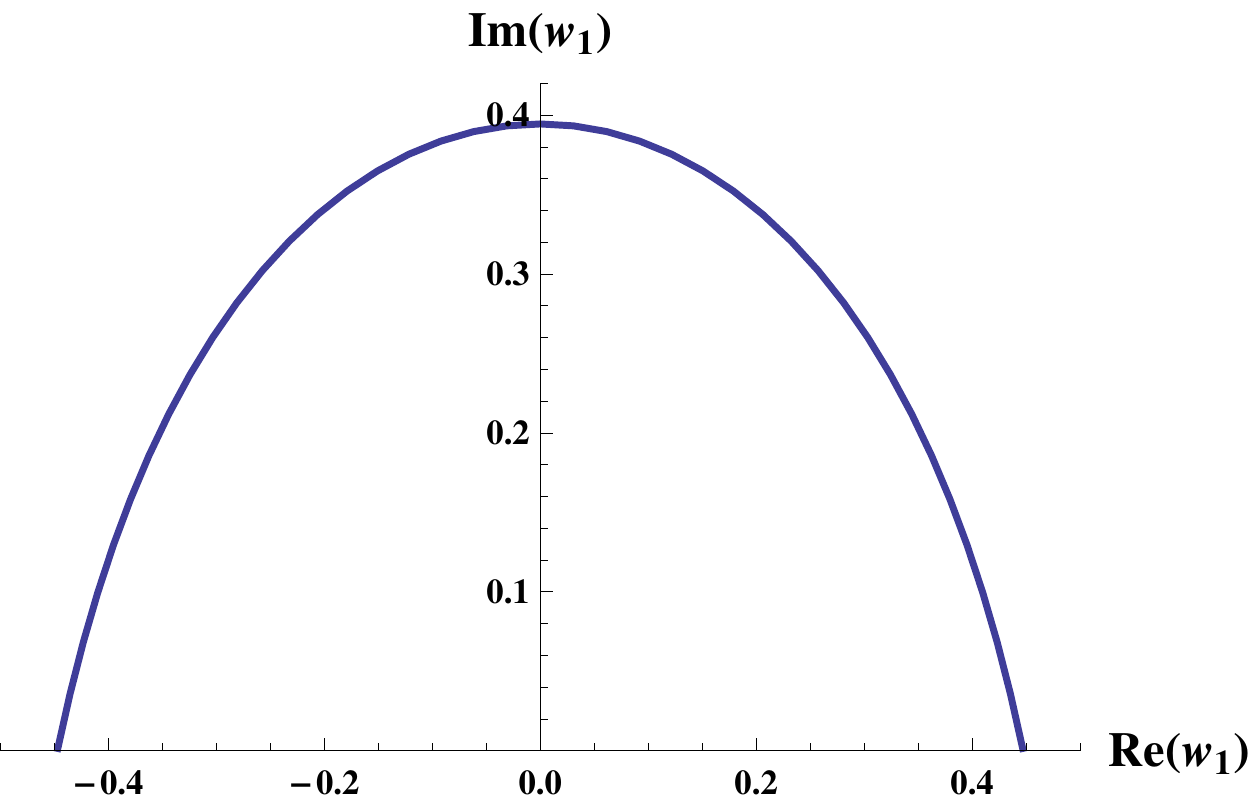}
  \hskip 0.5in
  \includegraphics[height=42mm]{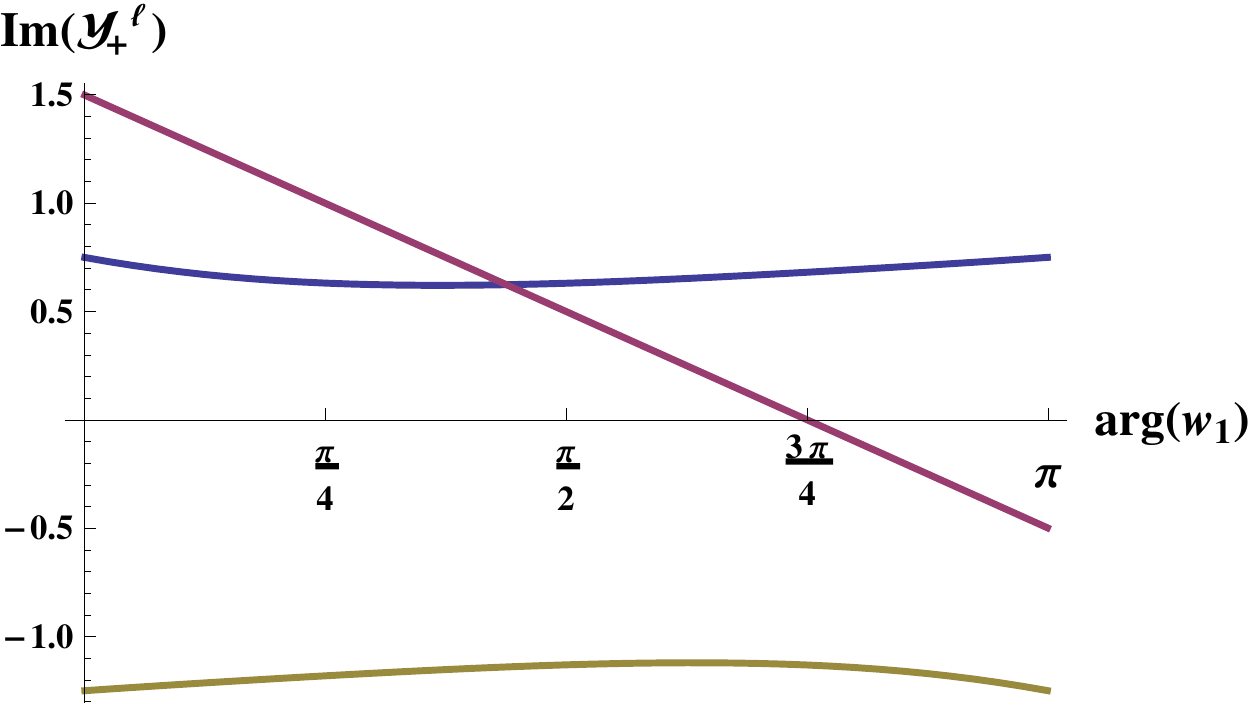}
  \caption{On the left hand side the allowed locations for the branch point $w_1$ in the upper half plane,
  for the solution (\ref{eq:3pole-ex}) with a single puncture corresponding to a D7-brane and $n_1=1$.
  On the right hand side the imaginary part of the charges along the curve shown on the left.
  At $\arg(w_1)=\pi$ the curves are, from top to bottom, $\Im(\cY_+^1)$, $\Im(\cY_+^2)$ and $\Im(\cY_+^3)$.
  \label{fig:w1}
 }
\end{figure}
For the particular solution (\ref{eq:3pole-ex}), the resulting curve to which $w_1$ is restricted is shown in fig.~\ref{fig:w1}.
It is not a half circle as in the previous example but of similar form.
The curve starts and ends on the real line, between the poles $p_2$, $p_3$ and $p_1$, $p_2$, respectively.
For any value of $w_1$ along the curve, with $\cA_\pm^0$ as described above, all regularity conditions in (\ref{eq:w1-summary}) and (\ref{eq:DeltaG0-summary}) are solved.
We note that there is no direction along which the branch point could be moved out of $\Sigma$ along its branch cut for this choice of $\gamma_1$.
The puncture is ``trapped'' inside $\Sigma$ in that sense.
As in the previous example, the real parts of the residues are constant along the curve, and given by
\begin{align}
\Re(\cY_+^1)&=1 & \Re(\cY_+^2)&=-2 & \Re(\cY_+^3)&=1
\end{align}
But the imaginary parts vary, as shown on the right had side in fig.~\ref{fig:w1}.

\sm

To  explicitly construct the supergravity fields for a set of parameters that solve the regularity conditions as above, we now have to construct the locally holomorphic functions $\cA_\pm$ and the composite quantities $\kappa^2$, $\cG$ explicitly. We do this numerically as follows.
Once the regularity conditions are solved it is straightforward to construct the differentials $\partial_w\cA_\pm$ via (\ref{eqn:dcA-summary}). Constructing the locally holomorphic functions $\cA_\pm$ themselves, however, already requires a more non-trivial integration than in the case without monodromy, as is evident from the expression in (\ref{eqn:cA-summary}).
From the functions $\cA_\pm$ we then have to construct the locally holomorphic function $\cB$ defined in (\ref{eq:comp}) by a further integration.
With these functions in hand one can then construct $\cG$ and $R$ in (\ref{eq:comp}) and from those the metric functions via (\ref{eq:f2f6rho}), the axion-dilaton scalar $B$ via (\ref{2a6}) and the gauge field via (\ref{2a7}).
To explicitly construct the supergravity fields we implement a two-step numerical integration procedure.
In a first step we construct $\cI$ defined in (\ref{eq:I}) and from that the locally holomorphic functions $\cA_\pm$ on a dense grid in the upper half plane.
Since the $\cA_\pm$ feed into the construction of $\cB$ via a further integration, they are needed with higher precision than the desired precision for the supergravity fields.
To accurately capture the rapidly varying behavior of $\cA_\pm$ around the poles on the real line and around the branch cuts, the grid in particular contains a large number of points around the poles and also a large number of points closely tracing the branch cuts.
The freedom in choosing the integration contour in (\ref{eqn:cA-summary}) (illustrated in fig.~\ref{fig:branchcut}) can be exploited to avoid rapidly varying regions for all other points.
In a second step we then determine $\cB$ by a further numerical integration. The grid can be chosen less dense but again contains a large number of points around the poles and branch cuts, to accurately capture the behavior there. Once these functions are determined it is then straightforward to compute the supergravity fields.

\begin{figure}
  \centering
  \begin{tabular}{lll}
    \includegraphics[width=59mm]{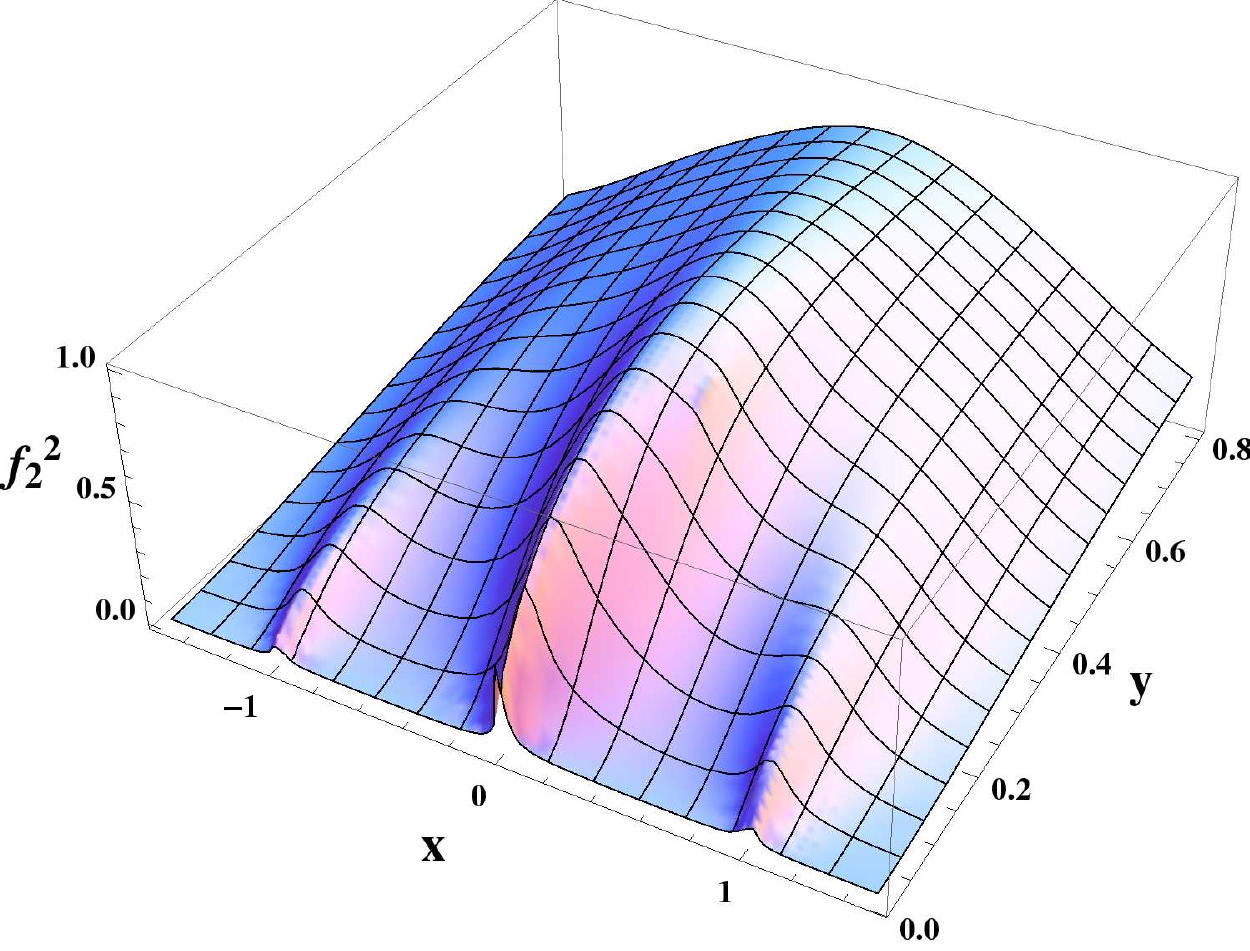}&\hskip 1in
    \includegraphics[width=59mm]{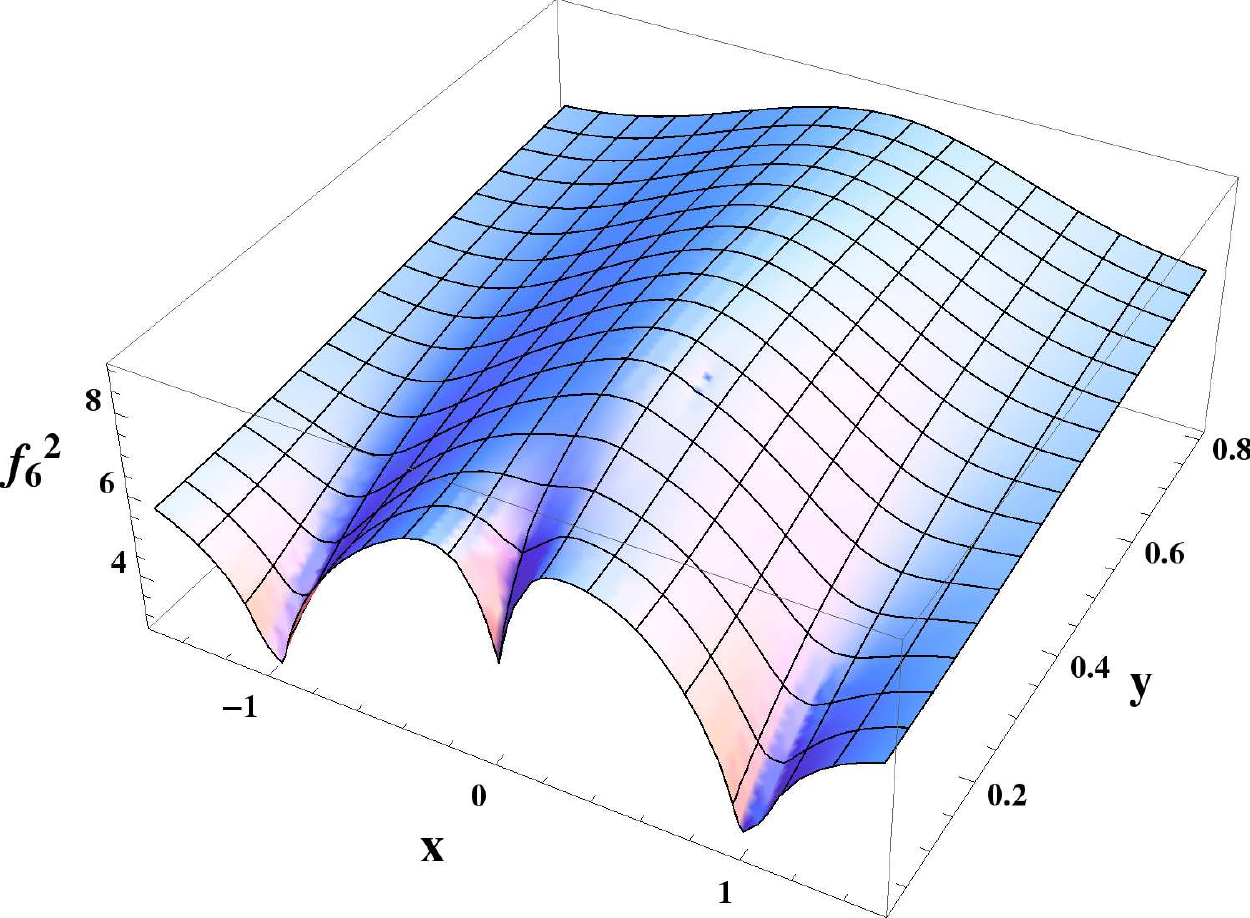}\\
    \includegraphics[width=59mm]{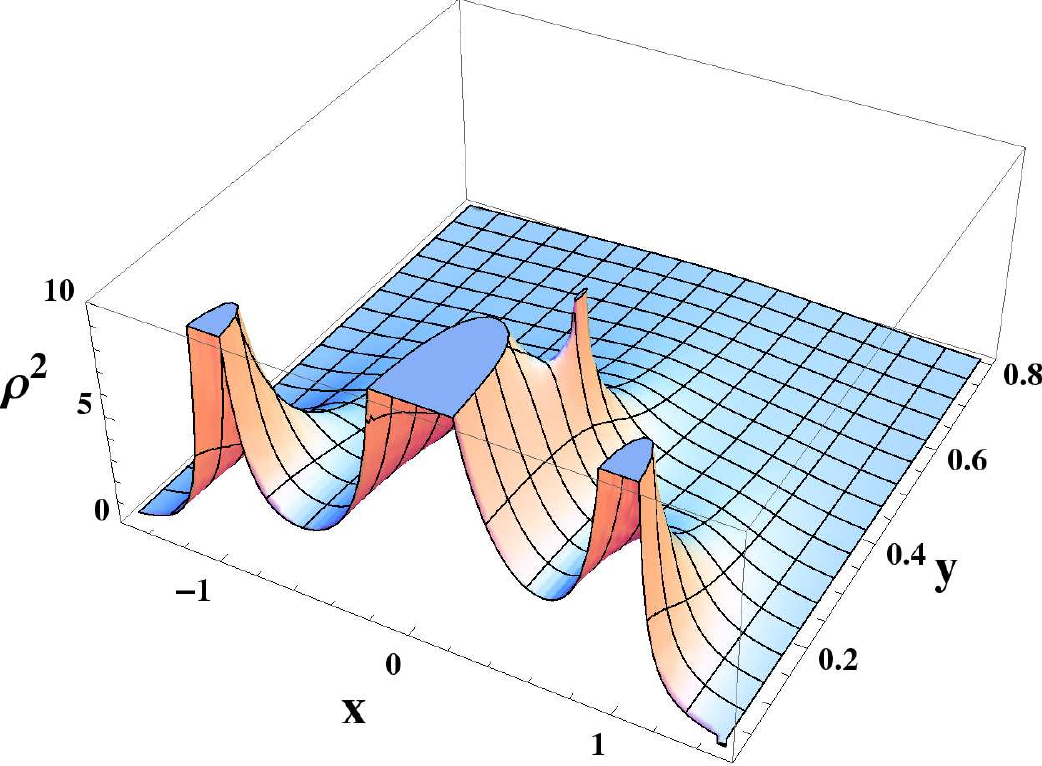}\\
    \includegraphics[width=59mm]{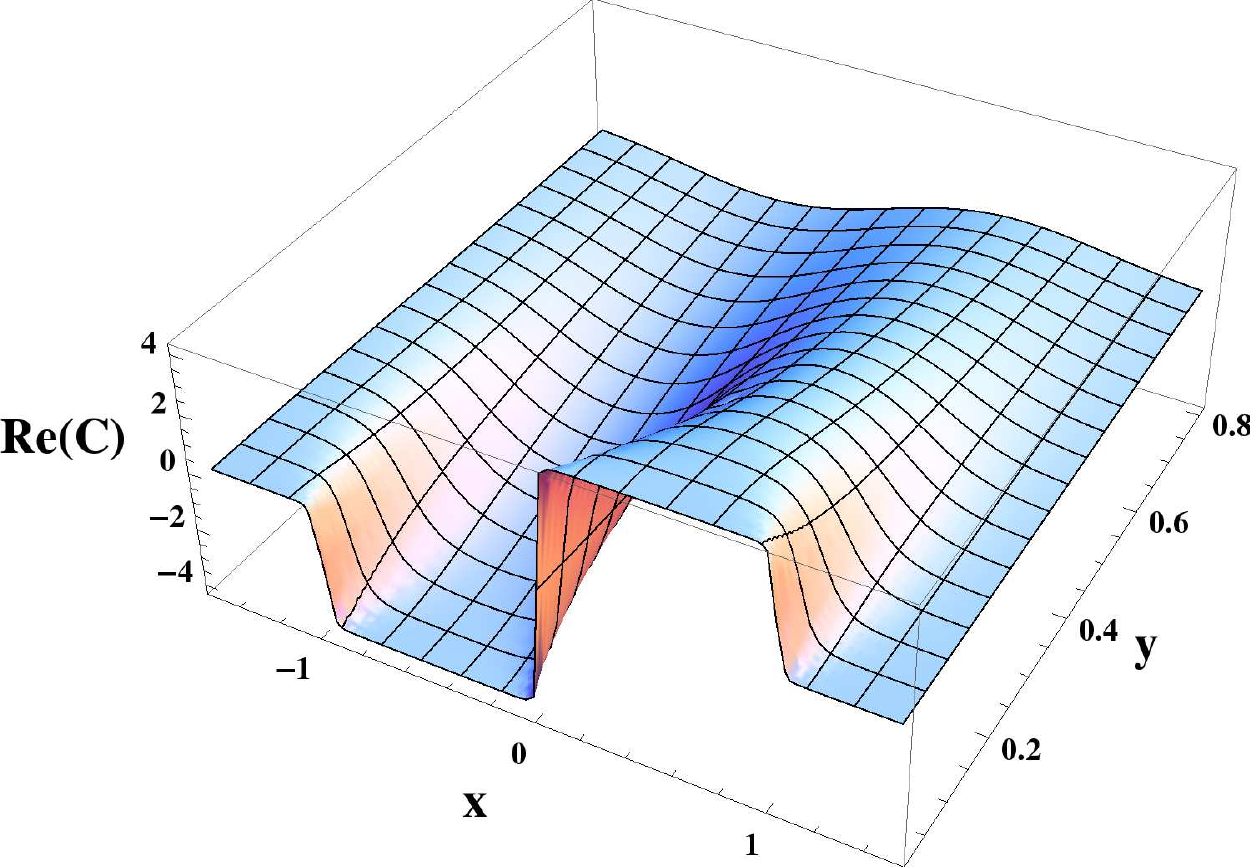}&\hskip 1in
    \includegraphics[width=59mm]{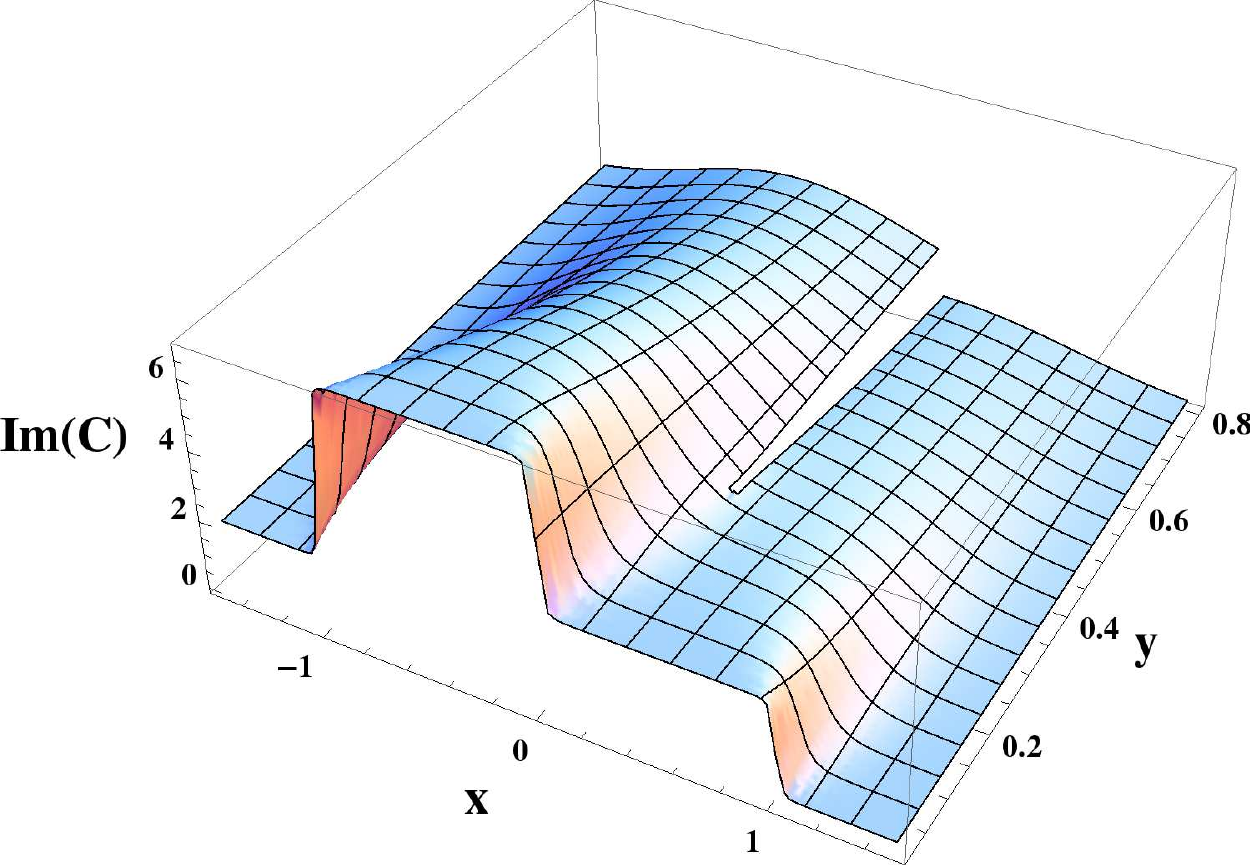}\\
    \includegraphics[width=59mm]{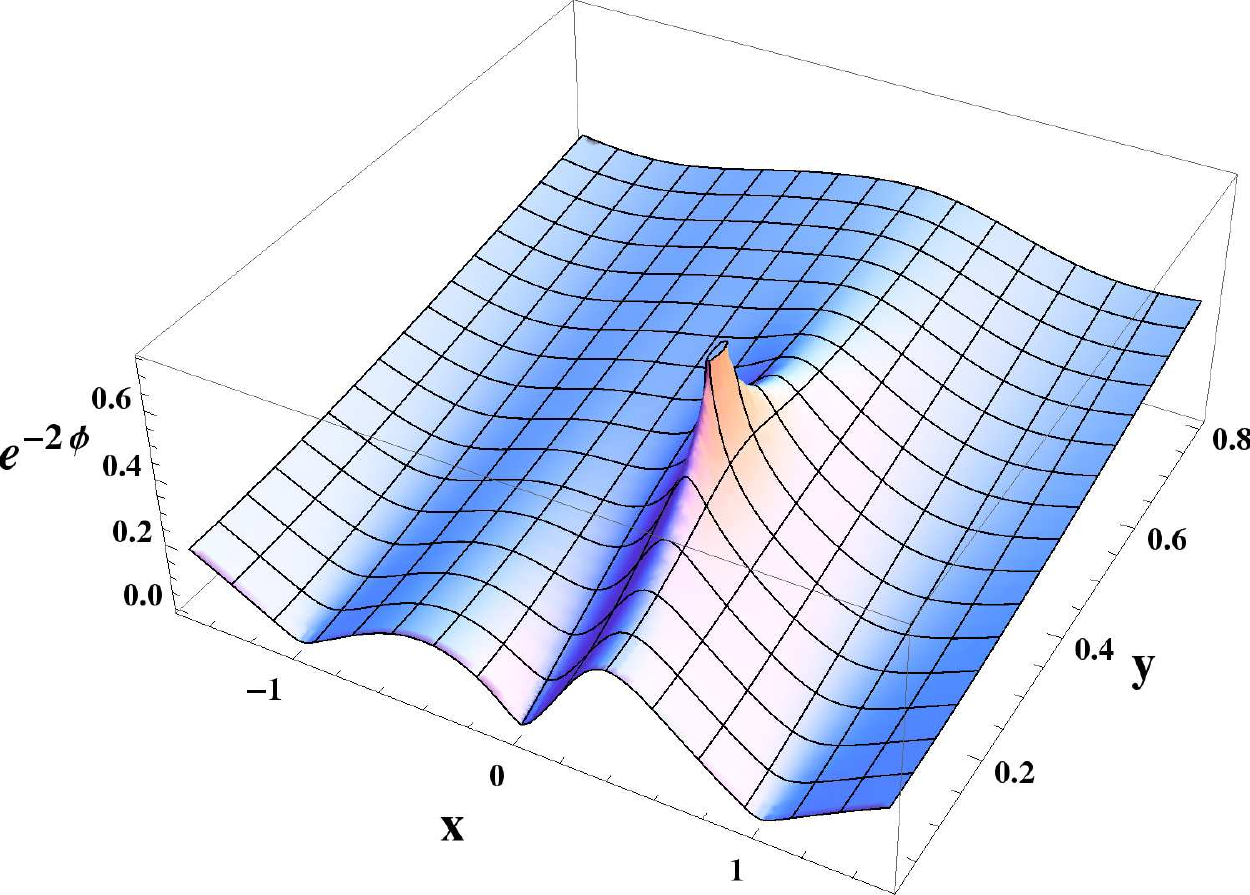}&\hskip 1in
    \includegraphics[width=59mm]{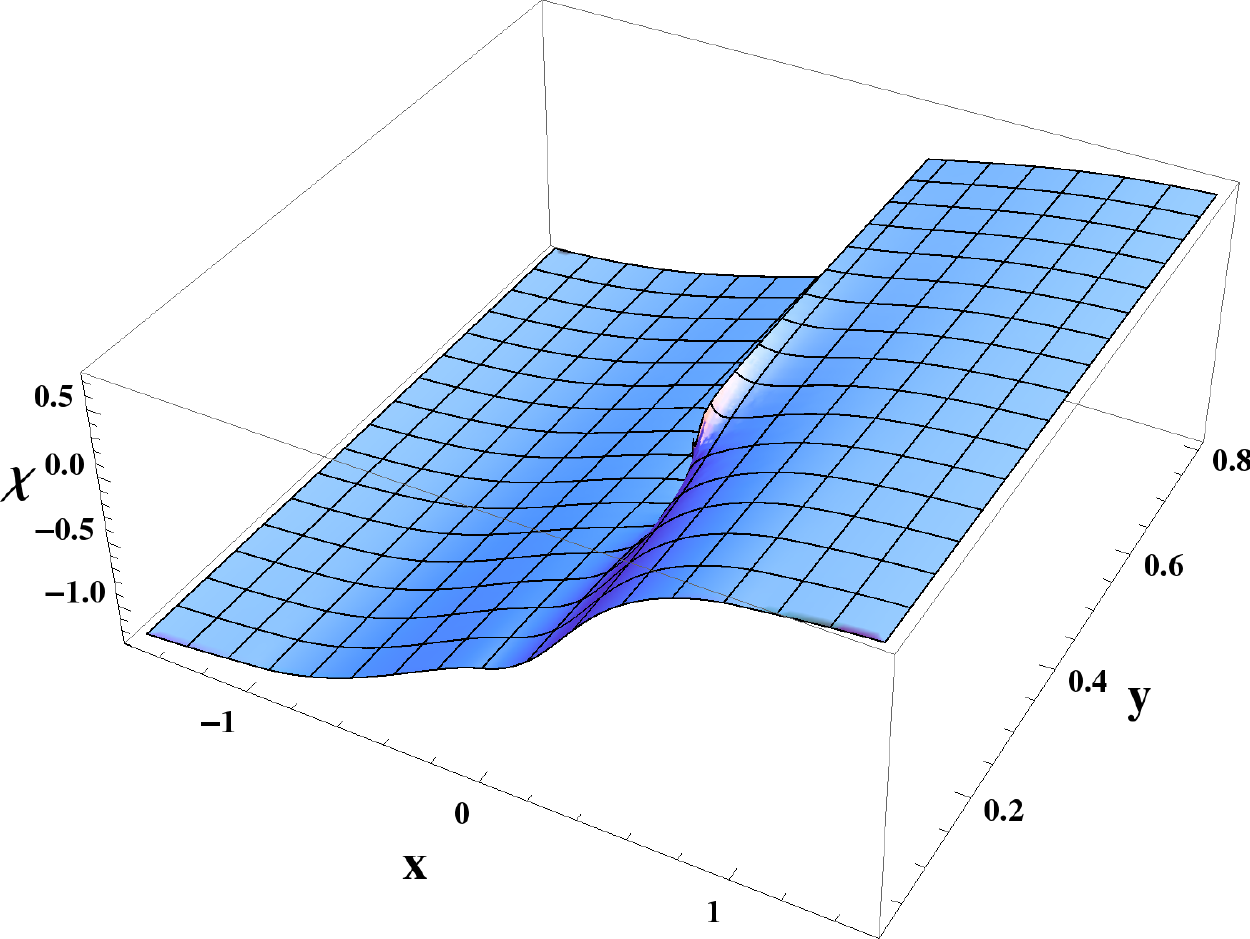}
  \end{tabular}
  \caption{
  The metric factors $f_2^2,f_6^2$ and $\rho^2$,
  the real and imaginary parts of the two-form potential $\cC$
  and axion and dilaton for the 3-pole solution with $[1,0]$ branch point.
  \label{fig:3-pole}
 }
\end{figure}

For the sake of presenting explicit plots of a solution, we pick a generic point on the curve shown in fig.~\ref{fig:w1}, namely
\begin{align}
w_1&=0.3980480542\,e^{i\pi/4}
\end{align}
Plots of the supergravity fields for the resulting solution are shown in fig.~\ref{fig:3-pole}.
They show that the branch cut indeed starts at $w_1$ and from there extends in the positive imaginary direction.
The plots also show that the metric functions are smooth and single-valued, with only $\rho^2$ diverging at the position of the D7-brane, as desired.
The dilaton blows up at the location of the D7-brane but is otherwise smooth, as expected, 
and the axion has non-trivial monodromy around $w_1$, realizing precisely the shift expected for a D7-brane.
The real part of the two-form field is smooth, and also the imaginary part behaves precisely as discussed in sec.~\ref{sec:asympt}.
Namely, $\cC$ transforms by the appropriate $SU(1,1)$ transformation combined with a constant gauge transformation such that the limit of $\cC$ as $w\rightarrow w_1$ is well defined.
The imaginary part of $\cC$ also reflects the fact that the imaginary parts of the residues in the presence of a D7-brane do not have to sum to zero: after crossing all three poles, the boundary value of $\Im(\cC)$ does not return to its original value.
The discrepancy in the value of $\Im(\cC)$ on the boundary to the left of all poles and to the right of all poles is given by the discontinuity of $\Im(\cC)$ across the branch cut at infinity.
The real parts of the residues, on the other hand, still sum to zero and correspondingly the value of $\Re(\cC)$ on $\partial\Sigma$ does return to its original value after crossing all three poles.
The behavior of all fields at the poles on the real line is as expected for an identification of the poles with 5-branes, in the same way as discussed in more detail in~\cite{DHoker:2017mds}.

\subsection{3-pole solution with \texorpdfstring{$[0,1]$}{[0,1]} branch point}
\label{sec:01ex}

\begin{figure}
  \centering
  \begin{tabular}{lll}
    \includegraphics[width=59mm]{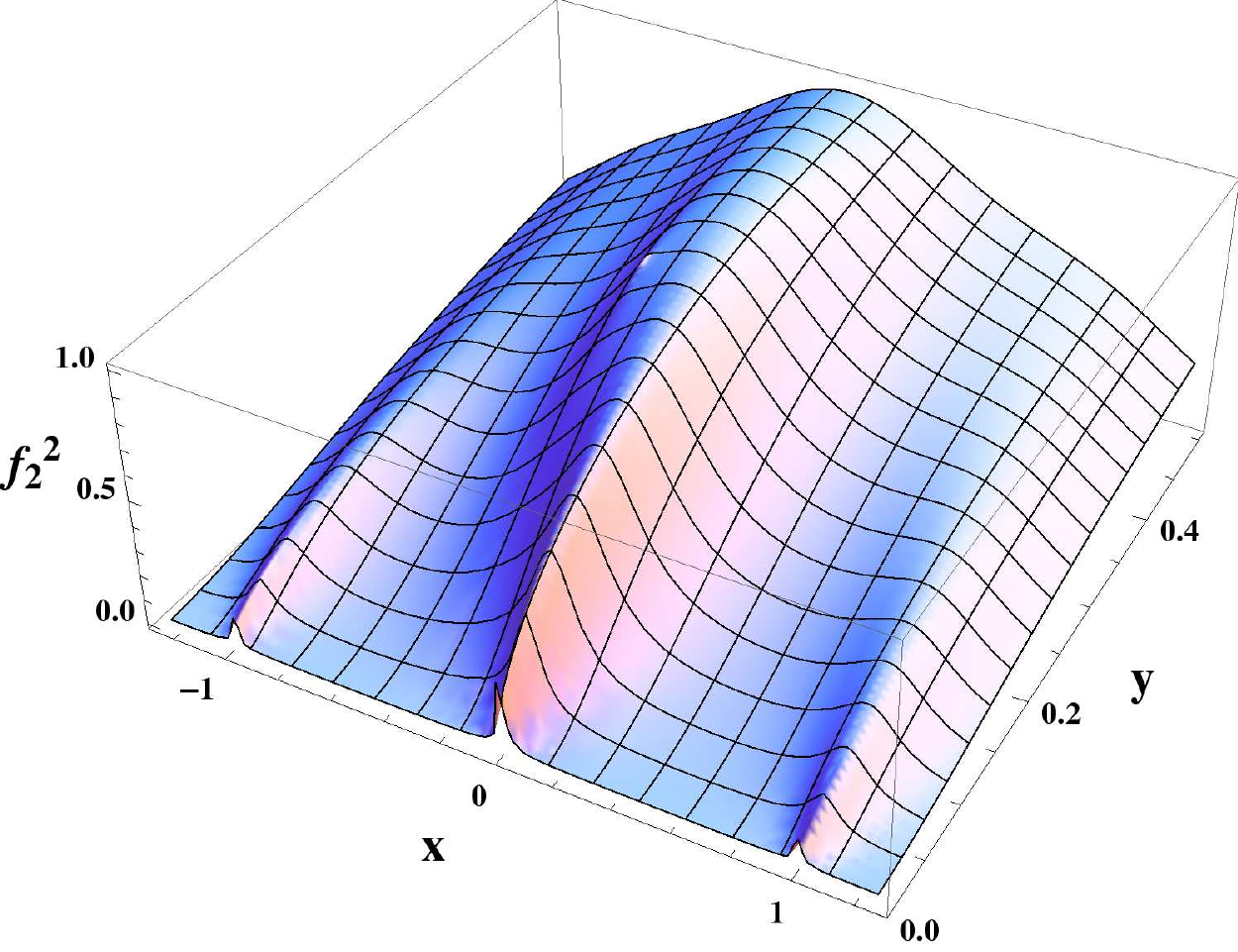}&\hskip 1in
    \includegraphics[width=59mm]{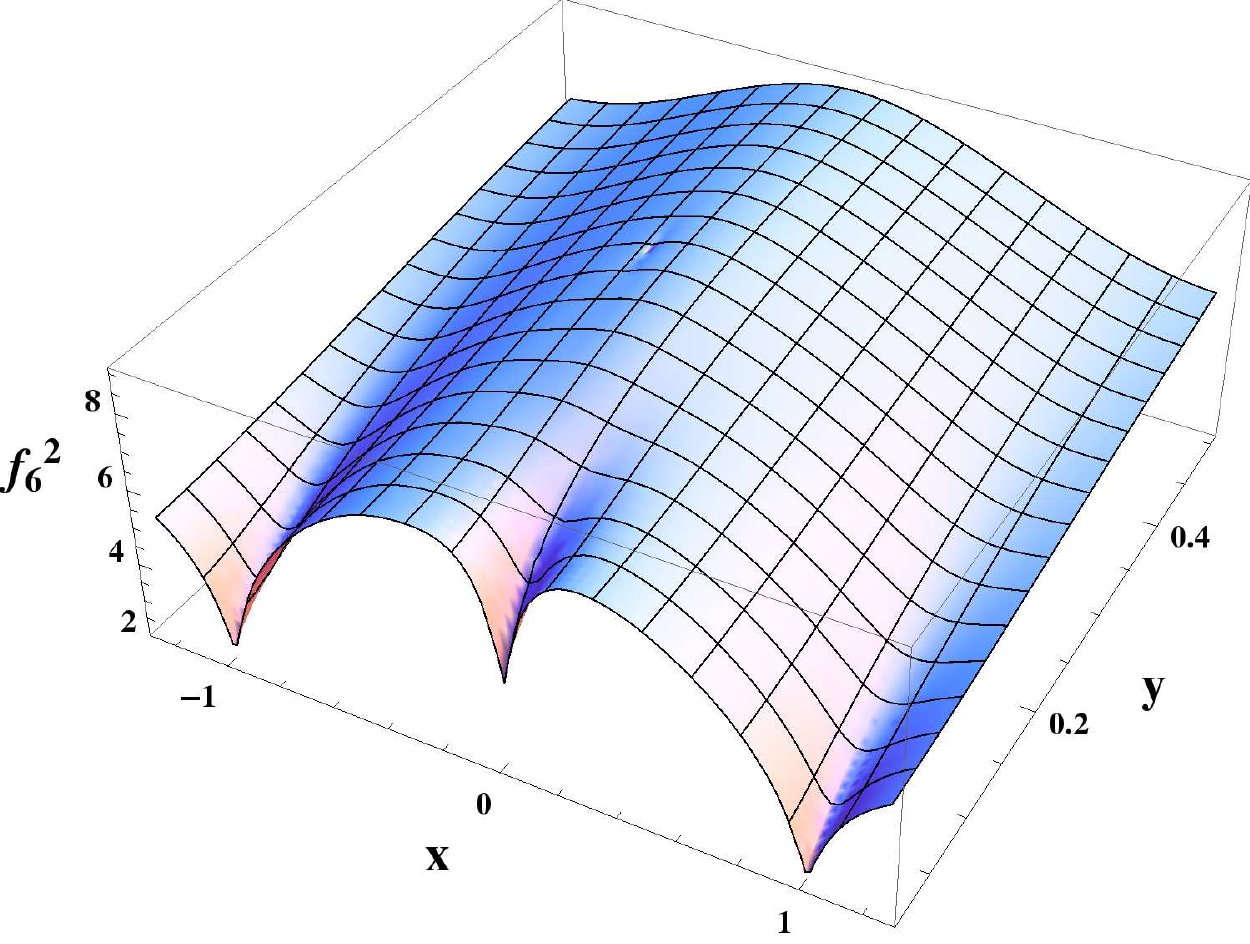}\\
    \includegraphics[width=59mm]{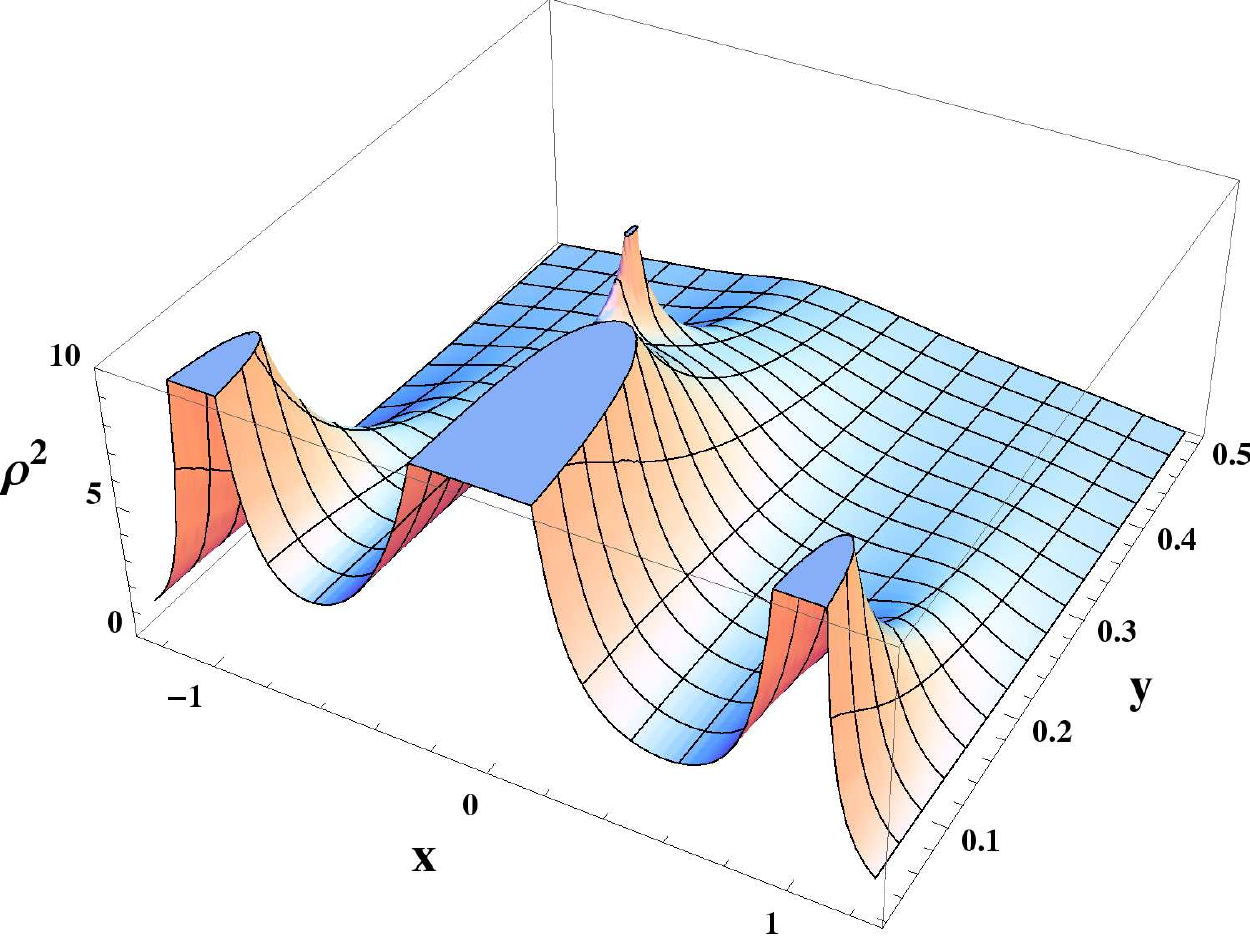}\\
    \includegraphics[width=59mm]{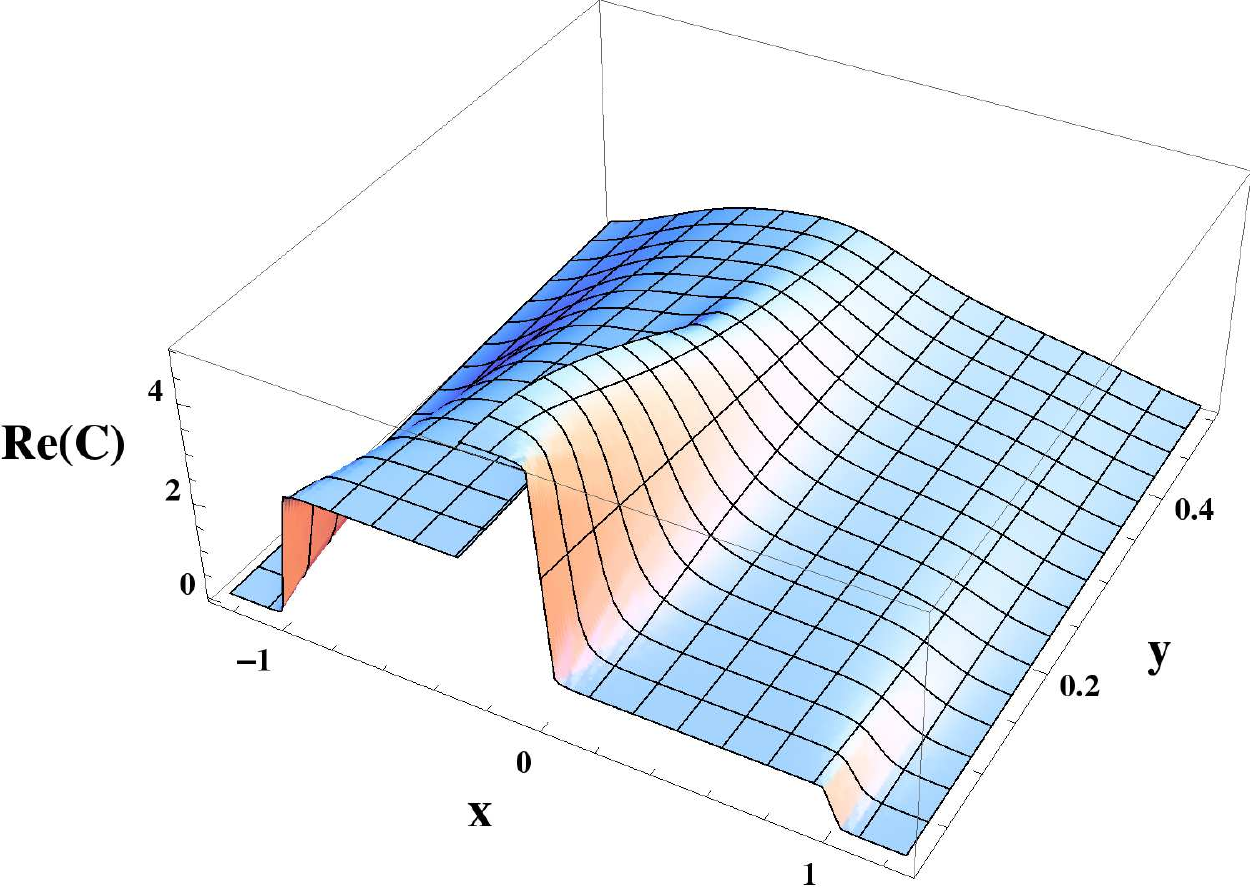}&\hskip 1in
    \includegraphics[width=59mm]{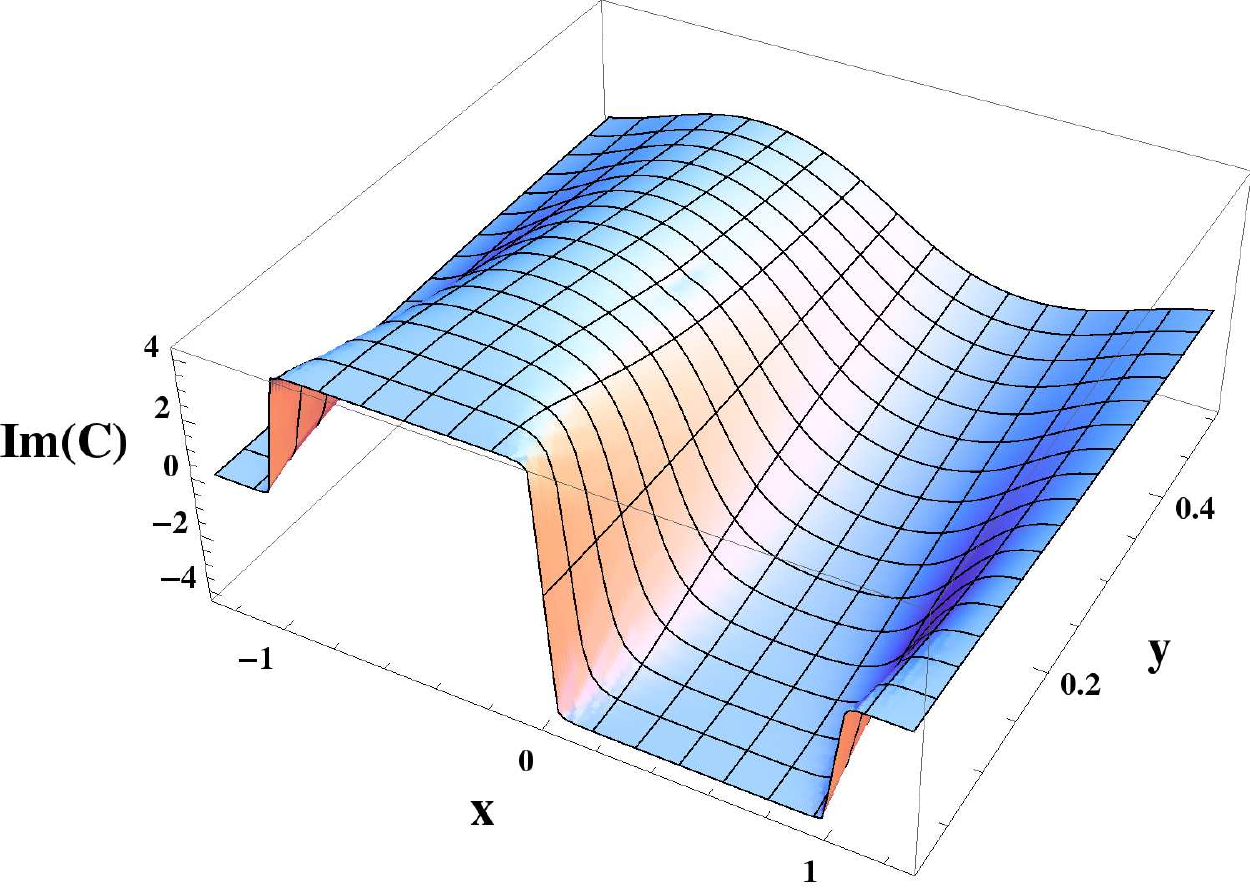}\\
    \includegraphics[width=59mm]{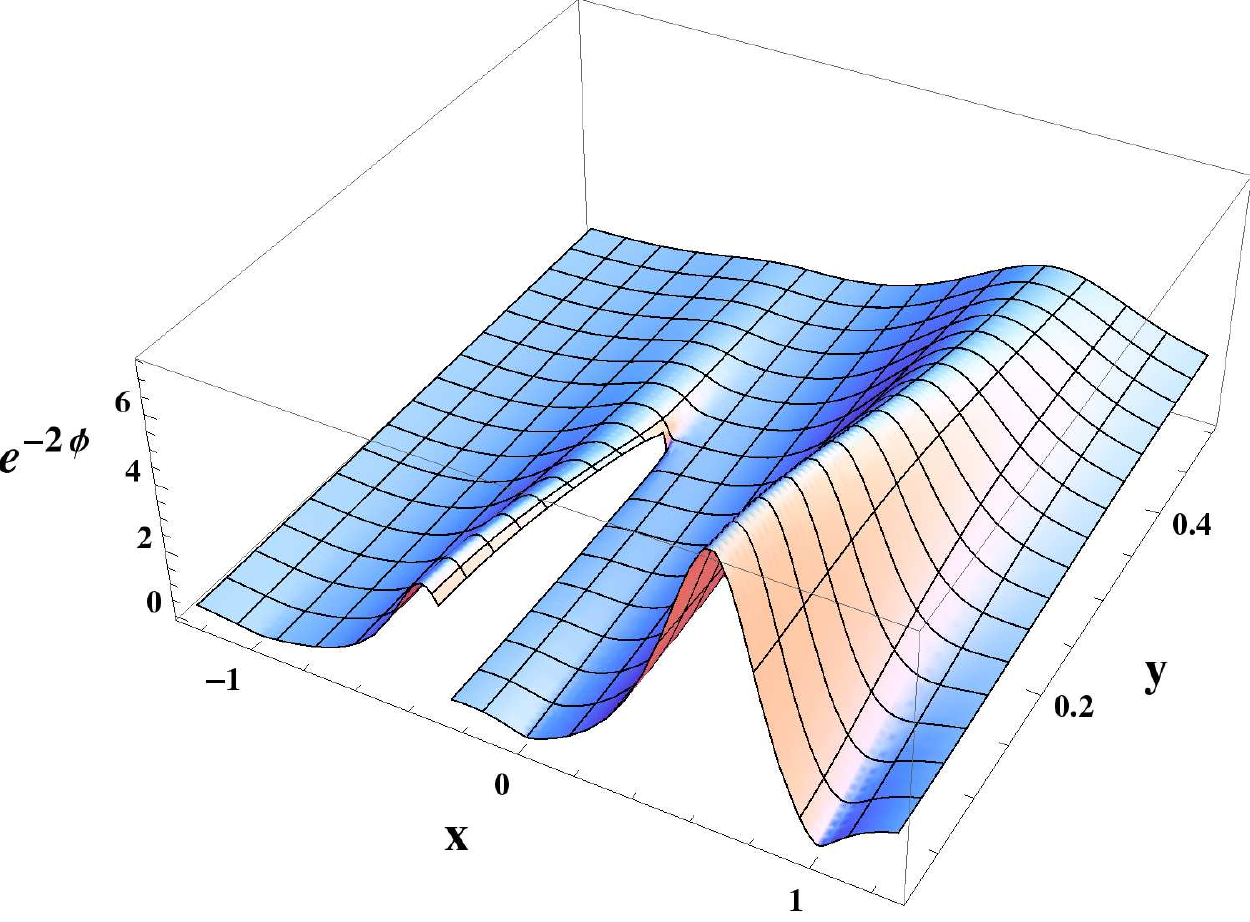}&\hskip 1in
    \includegraphics[width=59mm]{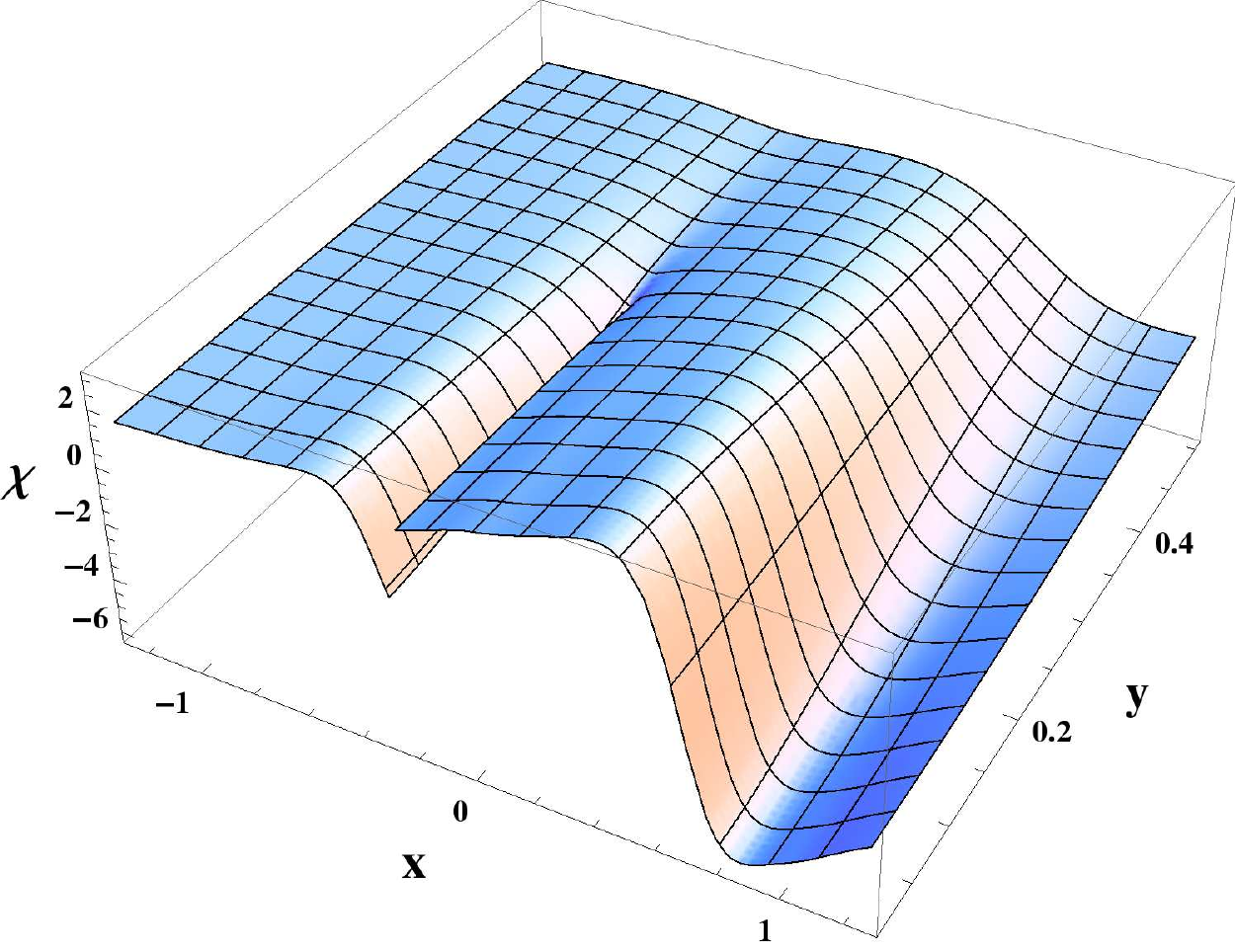}
  \end{tabular}
  \caption{
  The metric factors $f_2^2,f_6^2$ and $\rho^2$,
  the real and imaginary parts of the two-form potential $\cC$
  and axion and dilaton for the 3-pole solution with $[0,1]$ monodromy.
  \label{fig:3-pole-2cut}
 }
\end{figure}

As a second explicit example we will consider a case with a different choice of the charges and a different orientation of the branch cut, to illustrate the features of the solutions in that case. 
We start again from the 3-pole solution (\ref{eq:3-pole-pos}) with (\ref{eq:3pole-ex}), and add a branch point with $\pqseven{}=[0,1]$ monodromy, corresponding to the S-dual of a D7-brane.
Choosing $\pqseven{}=[0,1]$ results, via (\ref{eq:uqvq}), in $\eta_+=-\eta_-=u_Q=-i$ and $v_Q=0$, and we fix
\begin{align}
 n_1&=1& \gamma_1&=1
\end{align}
Solving the regularity conditions proceeds in the same way as outlined for the previous example, and the location of the branch point is once again restricted to a curve in $\Sigma$ which can be parametrized by $\arg(w_1)$.
From the expression for the residues at the poles on the real line in (\ref{eq:cY}) we now see that their imaginary part is unaffected by the addition of the branch point, but their real parts change.
The conserved linear combination of the \pq{} 5-brane charges therefore is the D5-charge, corresponding to the imaginary parts of the residues. The NS5-charge, corresponding to the real parts of the residues, is modified and in general not conserved.
To show explicit solutions we again pick a generic point on the curve, namely
\begin{align}
 w_1&=0.3980480542\, e^{3i\pi/4}
\end{align}
The residues for this particular choice for the location of the branch point are given by
\begin{align}
 \cY_+^1&=0.181179 -i
 &
 \cY_+^2&=1.000000+2i&
 \cY_+^3&=-0.631161-i
\end{align}
where the imaginary parts are exact and the real parts evidently do not sum to zero.
Plots of the metric functions, the two-form gauge field and the axion and dilaton for that solution are shown in fig.~\ref{fig:3-pole-2cut}.
The behavior of the metric functions is qualitatively similar to the example with $[1,0]$ monodromy: the radii of $AdS_6$ and $S^2$ are finite at the branch point while $\rho^2$ diverges, as expected.
For the two-form gauge field, on the other hand, the imaginary part is now continuous across the branch cut, while the real part is not. Their roles are thus switched compared to the previous example, as expected.
The non-conservation of the real part of the residues at the poles is reflected in the values of $\Re(\cC)$ on the boundary as well: since there is no pole or branch cut at infinity, the  boundary value of $\Re(\cC)$ to the left of all poles equals its boundary value to the right of all poles, but the non-conservation is manifest in the discontinuity at the point where the branch cut intersects the real line.
Axion and dilaton now both behave non-trivially when crossing the branch cut, reflecting the expected behavior for a $[1,0]$ monodromy.
Moreover, the exponentiated dilaton $e^{-2\phi}$ is finite at the branch point, instead of diverging as previously for the branch point corresponding to a D7-brane. This is the  expected behavior after performing an S-duality transformation and completes the discussion of all the non-trivial supergravity fields.
In summary, we find a solution that satisfies the physical regularity conditions and realizes the desired monodromy.

\sm

The behavior of the supergravity fields for generic \pqseven{}  7-brane charges is qualitatively similar and shows a combination of the features seen for the specific examples we discussed in detail. In general, the real and imaginary parts of $\cC$ both have a discontinuity across the branch cut, corresponding to the fact that the conserved linear combination of the charges does not simply reduce to the real or imaginary part of the residues.
Likewise, as seen already for the $[0,1]$ example, axion and dilaton both transform non-trivially. 
The exponentiated dilaton $e^{-2\phi}$ is finite at the branch point when $q\neq 0$ and diverges if $q=0$.
The generalization to multiple branch points with commuting monodromies is likewise straightforward, the plots become more busy but the regularity conditions derived in sec.~\ref{sec:sol} again guarantee smooth metric functions and that the two-form gauge field and the axion-dilaton scalar show the desired behavior across the branch cuts.

\newpage

\section{Connection to 5-brane webs with 7-branes}\label{sec:5-brane-webs}
\setcounter{equation}{0}

In this section we will discuss the connection of the supergravity solutions constructed in sec.~\ref{sec:sol} to 5-brane webs with additional 7-branes in more detail. 
We will first revisit the identification with 5-brane intersections and
then turn to the punctures and their identification with additional 7-branes.

\sm

As argued in \cite{DHoker:2017mds}, the solutions without monodromy have a compelling interpretation as supergravity descriptions for fully localized intersections of 5-branes, as obtained by taking the conformal limit of 5-brane webs describing 5d gauge theories. 
The arguments were based on having the correct symmetries and parameter count, and in particular on the identification of the poles on the real line with the external 5-branes defining the intersection.
This identification directly carries over to the solutions with monodromy, since it only uses the leading behavior of the holomorphic data close to the poles and the differentials for the solutions with monodromy again have simple poles on the real line.
By direct extension of the identification in sec.~\ref{sec:poles-5branes}, we therefore find that the poles $p_\ell$ on the real line correspond to 5-branes with charges determined by the residues $\cY_\pm^\ell$ in (\ref{eq:cY}).
Analogously to (\ref{eq:Zp-charge}), the identification with the charge vector $(q_1,q_2)Q$ in the conventions of \cite{Lu:1998vh} is given by
\begin{align}
 (q_1-iq_2)Q&=\frac{8}{3}c_6^2\cY_+^\ell
\end{align}
with the real part of $\cY_+^\ell$ corresponding to NS5 charge and the imaginary part corresponding to D5 charge.
Compared to the $Z_\pm^\ell$ which determined the charges in the solutions without monodromy, however, the residues $\cY_\pm^\ell$ are less constrained.
For solutions with D7-branes, only the real parts of the $\cY_\pm^\ell$ have to sum to zero: 
Since $f(p_\ell)$ is imaginary,  $\eta_\pm=1$ and $Y^\ell$ real, eq.~(\ref{eq:cY}) shows that the real parts of $\cY_\pm^\ell$ sum to zero, due to charge conservation in the seed solution without monodromy. But the imaginary parts  in general do not.
This was clearly exhibited in the example solutions discussed in sec.~\ref{sec:3-pole-ex-2} and \ref{sec:3pole-1cut}, where the sum over the imaginary parts of the residues was non-vanishing.
For general \pqseven{} 7-branes the corresponding $SL(2,\RR)$ rotated statements hold, and we likewise have one real charge conservation constraint on the complex residues.  For $[0,1]$ 7-branes this simply corresponds to switched roles for the real and imaginary parts of the residues, as exhibited in the example in sec.~\ref{sec:01ex}. We therefore find that the solutions correspond, in general, to 5-brane intersections with only one linear combination of the \pq{} 5-brane charges conserved.

\sm

We now come to the punctures themselves.
The parabolic $SL(2,\RR)$ monodromies given in (\ref{eq:Mpq}) have the expected form for a \pqseven{} 7-brane \cite{Gaberdiel:1997ud}, and for multiple coincident branes we expect precisely a monodromy of the form given in (\ref{eq:Mpq-n}).
As discussed in sec.~\ref{sec:d7} the punctures can indeed be identified with \pqseven{} 7-branes, and as reviewed in the introduction the addition of 7-branes into 5-brane webs is well motivated.
The way they appear in our solutions indeed matches well with their role in the 5-brane webs.
To recall, if we take the 5-branes in the string theory construction to extend along the directions $0-4$ and a one-dimensional subspace of the $5-6$ plane, then the  7-branes are localized at points in the $5-6$ plane and wrap all other directions, as summarized in the following table \cite{DeWolfe:1999hj}:
\vskip 2mm
\begin{center}
\begin{tabular}{c||cccccccccc}
\hline
\hline
& 0&1&2&3&4&5&6&7&8&9\\
\hline
\hline
 D5 brane & x & x & x & x & x & x\\
 NS5 brane & x & x & x & x & x &  & x \\
 7-brane & x & x & x & x & x & & & x & x & x\\
\hline
\hline
\end{tabular}
\end{center}
\vskip 2mm
In our supergravity solutions the poles on the boundary of $\Sigma$ represent the remnants of the semi-infinite external 5-branes, which suggests that $\Sigma$ encodes the structure of the web in the $5-6$ plane.
 We would then expect each 7-brane to be localized at a point in $\Sigma$ and wrap all other parts of the geometry,
precisely as we find from the discussion in sec.~\ref{sec:d7}.
The fact that we naturally found D7-branes and their $SL(2,\RR)$ orbits of \pqseven{} 7-branes in sec.~\ref{sec:d7}, instead of anti D7-branes, also has a natural interpretation from the brane web perspective.
While for a 7-brane alone both choices are possible and supersymmetric, the difference becomes crucial in the presence of the 5-branes. 
To preserve supersymmetry, the 7-branes added to a 5-brane web have to be compatible with precisely the supersymmetries preserved by the 5-branes, hence explaining the restriction to D7-branes and their $SL(2,\RR)$ orbits.
The presence of 7-branes also provides a natural brane web explanation for the fact that the residues $\cY_\pm^\ell$, corresponding to the charges of the external 5-branes, do not necessarily sum to zero, as discussed in the previous paragraph. 5-branes may cross the branch cuts introduced by the 7-branes, where their charges undergo the corresponding $SL(2,\RR)$ transformation and thus potentially change. Moreover, 5-branes can terminate on the 7-branes, such that their charges do not contribute to the total charge of the external 5-branes at all. The total charges of the external 5-branes therefore do not necessarily sum to zero in the presence of 7-branes, precisely as realized in the supergravity solutions.
We thus find a coherent general picture where the supergravity solutions constructed in sec.~\ref{sec:sol} correspond to the conformal limit of 5-brane webs with additional 7-branes.

\sm

Establishing a precise map between specific brane webs and our supergravity solutions is beyond the scope of this work, but we will close this section with a speculative general discussion of a possible relation.
Since the supergravity solutions correspond to the conformal limit of 5-brane webs and the 7-branes are accessible in the supergravity description, a natural possibility would be that the solutions correspond to 5-brane webs with 7-branes inside the faces of the web.
This interpretation aligns well with the fact that we find 7-branes in a non-trivial background, as discussed in sec.~\ref{sec:d7}:
Taking the conformal limit of a 5-brane web with a 7-brane kept inside a face means the 7-brane ends up precisely on the 5-brane intersection.
The geometry created by the 5-branes at that point is $AdS_6\times S^2$ warped over $\Sigma$, and we thus find the 7-brane wrapping $AdS_6\times S^2$. There is no limit of moving along the 7-brane in the $5,6$ directions which would take us away from the intersection, such that we would expect to recover a 7-brane in flat space.
This is in contrast to the external 5-branes, where we can move along their worldvolume away from the intersection in the $5,6$ directions,
and gives a brane web interpretation for the discussion in sec.~\ref{sec:match}.
\begin{figure}[htb]
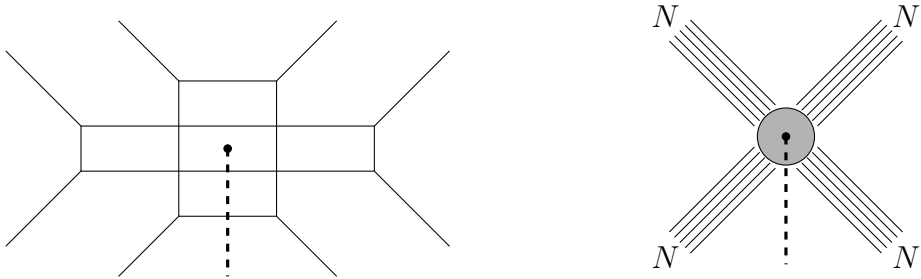

\begin{center}
\tikzpicture

\draw (-0.5,0.3) -- (3.4,0.3) -- (3.4,-0.3) -- (-0.5,-0.3) -- (-0.5,0.3);
\draw (0.8,0.9) -- (2.1,0.9) -- (2.1,-0.9) -- (0.8,-0.9) -- (0.8,0.9);

\draw (-0.5,0.3) -- (-0.5-1,0.3+1);
\draw (-0.5,-0.3) -- (-0.5-1,-0.3-1);

\draw (0.8,0.9) -- (0.8-0.8,0.9+0.8);
\draw (2.1,0.9) -- (2.1+0.8,0.9+0.8);

\draw (3.4,0.3) -- (3.4+1,0.3+1);
\draw (3.4,-0.3) -- (3.4+1,-0.3-1);

\draw (0.8,-0.9) -- (0.8-0.8,-0.9-0.8);
\draw (2.1,-0.9) -- (2.1+0.8,-0.9-0.8);

\draw[fill=black] (1.45,0) circle (0.05);
\draw[dashed,very thick] (1.45,0) -- (1.45,-1.7);

\endtikzpicture
\hskip 1in
\tikzpicture
\draw (0.28,0.28) -- (1.41,1.41);
\draw (0.35,0.21) -- (1.48,1.34);
\draw (0.42,0.14) -- (1.55,1.27);
\draw (0.21,0.35) -- (1.34,1.48);
\draw (0.14,0.42) -- (1.27,1.55);

\draw[fill=gray!60] (0,0) circle (0.38);

\draw [fill=black] (0,0) circle (0.05);
\draw[dashed,very thick] (0,0) -- (0,-1.7);

\draw (0.28,-0.28) -- (1.41,-1.41);
\draw (0.35,-0.21) -- (1.48,-1.34);
\draw (0.42,-0.14) -- (1.55,-1.27);
\draw (0.21,-0.35) -- (1.34,-1.48);
\draw (0.14,-0.42) -- (1.27,-1.55);

\draw (-0.28,-0.28) -- (-1.41,-1.41);
\draw (-0.35,-0.21) -- (-1.48,-1.34);
\draw (-0.42,-0.14) -- (-1.55,-1.27);
\draw (-0.21,-0.35) -- (-1.34,-1.48);
\draw (-0.14,-0.42) -- (-1.27,-1.55);

\draw (-0.28,0.28) -- (-1.41,1.41);
\draw (-0.35,0.21) -- (-1.48,1.34);
\draw (-0.42,0.14) -- (-1.55,1.27);
\draw (-0.21,0.35) -- (-1.34,1.48);
\draw (-0.14,0.42) -- (-1.27,1.55);

\node at (-1.6,-1.6) {$N$};
\node at (-1.6,1.6) {$N$};
\node at (1.6,-1.6) {$N$};
\node at (1.6,1.6) {$N$};

\endtikzpicture

\end{center}
\caption{
Brane web and intersection with a large-$N$ limit. 
On the right hand side the conformal limit for generic $N$, 
on the left hand side for $N=2$ a deformation corresponding to finite gauge coupling and a state on the Coulomb branch.
\label{fig:large-N-web}
}
\end{figure}
One might wonder in that context what the modulus corresponding to the position of the 7-brane in $\Sigma$, as exhibited in the parameter count in sec.~\ref{sec:sol} and in the examples in sec.~\ref{sec:examples}, would correspond to in the brane web picture when the 7-brane is trapped at the intersection point.
An explanation can be given by the fact that we are considering solutions corresponding to brane webs in a ``large-N'' limit.
Such brane webs can have a complex internal structure, as illustrated for an example in fig.~\ref{fig:large-N-web}.
The web for $N=2$ has four distinct faces, and in the limit where the charges of the external branes are large this becomes a dense grid of faces in which we can have a 7-brane.
The discrete choice of which face the 7-brane is in remains in the conformal limit where the web collapses to an intersection,
and in the large-$N$ limit it becomes effectively continuous.
In our supergravity solutions we expect the internal structure of the web to be encoded in $\Sigma$, 
and the choice of position of the branch point could then naturally correspond to the choice of face in which the 7-brane is located.
A similar argument can explain the choice for the orientation of the branch cuts, determined by the continuous parameters $\gamma_i$.
The trajectories of the branch cuts in $\Sigma$ could have a natural interpretation as corresponding to their trajectory through the dense grid of faces in the corresponding brane webs in the large-$N$ limit. 
This choice again remains meaningful in the conformal limit, giving a possible interpretation for all additional parameters associated with the punctures.

\newpage

\section{Discussion}
\setcounter{equation}{0}
\label{sec:discussion}

We have constructed physically regular $AdS_6$ solutions to Type IIB supergravity with 16 supercharges, that realize the unique five-dimensional superconformal algebra $F(4)$ geometrically.  Similarly to the solutions in \cite{DHoker:2016ysh,DHoker:2017mds}, the geometry takes the form $AdS_6\times S^2$ warped over a two-dimensional Riemann surface $\Sigma$. Moreover, there are once again mild isolated singularities on the boundary of $\Sigma$ that correspond to semi-infinite 5-branes. 
The new feature compared to the existing solutions is that $\Sigma$ has punctures around which the supergravity fields undergo non-trivial $SL(2,\RR)$ monodromy. The solutions may in that sense also be regarded as solutions to F-theory \cite{Vafa:1996xn}.\footnote{In the context of AdS$_3$/CFT$_2$, solutions with non-trivial monodromy were recently constructed in~\cite{Couzens:2017way}.}
We have identified the punctures with \pqseven{} 7-branes, and the fact that we can identify both, 5-branes and 7-branes, suggests a direct identification of the solutions with the conformal limit of 5-brane webs with additional 7-branes, as introduced in \cite{DeWolfe:1999hj}.
The solutions therefore provide compelling candidates for holographic duals of the UV fixed points of five-dimensional gauge theories that are described by brane webs with additional 7-branes.
This offers a clear path for quantitative analyses of the UV fixed points, 
e.g.\ of their spectra, entanglement entropies and free energies.
We will close with a discussion of open questions and of some directions for future research.

\sm

We have collected a number of arguments for the identification of the punctures with 7-branes already, and found a coherent general picture for the interpretation of the solutions we have constructed.
To further specify and substantiate the relation to 5-brane webs with additional 7-branes, a natural next step is to compare supergravity computations, e.g.\ of the free energy, to the corresponding field theory or string theory calculations.
Moreover, for the identification of the punctures with the addition of 7-branes additional consistency checks can already be performed directly in the supergravity description.
Namely, via the relation of 5-brane webs with 7-branes to 5-brane webs without 7-branes by the Hanany-Witten brane creation effect  \cite{DeWolfe:1999hj}. 
It suggests that certain supergravity solutions with punctures, as constructed here, should yield equivalent results in holographic computations as certain solutions without 7-branes, as constructed previously in \cite{DHoker:2016ysh,DHoker:2017mds}.
Identifying precisely which solutions are equivalent in that sense would provide interesting information about the internal structure of the webs and further support the identification of the supergravity solutions with brane webs.
A more technical question in that context concerns the role of the punctures for holographic computations: As shown in \cite{Gutperle:2017tjo}, the isolated singularities on the real line do not interfere with supergravity computations at least of the free energy and entanglement entropy. We expect the same to be true for the punctures since the singularities are of a similarly mild type, but leave an explicit verification for the future.

\sm

Concerning the solutions themselves, a natural next question is for an extension of the constructions presented here to include punctures with non-commuting monodromies. 
We have currently allowed for an arbitrary number of punctures with the restriction that the associated monodromies commute, which realizes mutually local 7-branes.
But in the brane web constructions mutually non-local 7-branes and the corresponding branch cut moves also play a prominent role, and it would therefore be desirable to have supergravity solutions with the corresponding features.

\section*{Acknowledgements}

We are happy to thank Oren Bergman, Andreas Karch and Diego Rodriguez-Gomez for many interesting discussions on five-brane webs.
We also acknowledge the Aspen Center for Physics, which is supported by National Science Foundation grant PHY-1066293, for hospitality during the workshop ``Superconformal Field Theories in $d\geq 4$'' and thank the organizers and participants for the enjoyable and inspiring conference.
ED is grateful to the Kavli Institute for Theoretical Physics in Santa Barbara for their hospitality during the completion of this work.
The work of all three authors is supported in part by the National Science Foundation under grant PHY-16-19926.
The work of ED is also supported in part by the National Science Foundation under grant NSF PHY-1125915.

\newpage

\appendix

\section{The vanishing of  \texorpdfstring{$\cG$ on $\partial\Sigma$}{G=0 on dSigma}}\label{app:G0}

In this appendix we provide further technical details for the derivation of the regularity conditions to guarantee $\cG=0$ in sec.~\ref{sec:DeltaG}. There are two auxiliary results for which we omitted the derivation in the main part and we will discuss the details in the following. 

\sm

The first result used in sec.~\ref{sec:DeltaG} is that the $\Delta_k\cB$ contribution in (\ref{eq:deltaG-0}) indeed reproduces the first term, and to evaluate the result more explicitly to arrive at (\ref{eq:DeltaG0b}).
Evaluating the first term in (\ref{eq:deltaG-0}) explicitly, using (\ref{eq:etaY}), yields
\begin{align}
 \Delta_k\cG&=
 2\pi i\Big(\bar \cA^0 Y_+^k+\cA^0 Y_-^k+\sum_{\ell\neq k} Y^{[\ell, k]}\ln |p_\ell-p_k|\Big)
 +i\pi Y^k\left(\overline{\cI(p_k+\epsilon)}-\cI(p_k+\ep)\right)
 \nonumber\\&\hphantom{=}
 +i\pi f(p_k) Y^k\left(\eta_-\cA_+^s(p_k+\ep)-\overline{\eta_+\cA_-^s(p_k+\ep)}+\mathrm{c.c.}\right)
 +\Delta_k\cB+\Delta_k\bar\cB
 \label{eq:DeltaG0}
\end{align}
It remains to evaluate $\Delta_k\cB$. 
Starting from (\ref{eqn:DeltaB}) and using (\ref{eq:dA-split}), (\ref{eq:Asplit}) we find
\be\label{eq:DeltaB}
 \Delta_k\cB=\Delta_k\cB^s -  \int_{C_k} dz \, \cI\,(\eta_-\partial_z\cA_+^s-\eta_+\partial_z\cA_-^s)
 +\int_{C_k}dz \, \cF\,(\eta_-\cA_+^s-\eta_+\cA_-^s)
\ee
where $\cB^s$ denotes the part of $\cB$ constructed from the single-valued differentials and functions,
and $C_k$ is the half circle contour centered on $p_k$.
It is convenient to evaluate $\Delta_k\cB$ together with its complex conjugate.
For the last term we find
\be
 \int_{C_k}\!dz \, \cF(\eta_-\cA_+^s-\eta_+\cA_-^s)+\mathrm{c.c.}=
 i\pi f(p_k)Y^k\Big(\eta_-\cA_+^s(p_k+\ep)-\eta_+\cA_-^s(p_k+\ep)+\mathrm{c.c.}\Big)
 \hskip 0.2in
 \label{eq:DeltaB1}
\ee
where $\epsilon>0$ once again is the radius of the half circle $C_k$. 
For the second term in (\ref{eq:DeltaB}) we have to evaluate the integral in $\cI$ from $\infty$ to $z\in C_k$.
It is convenient to split it into the part from $\infty$ to the starting point of $C_k$, $p_k+\epsilon$, and the remaining part along the half circle $C_k$, parametrized by $p_k+\epsilon e^{i\theta}$ with $\theta \in (0,\pi)$. 
Namely,
\be
\label{eq:Iexpl}
 \cI(p_k+\epsilon e^{i\theta})=\cI(p_k+\epsilon)+i\epsilon\int_0^\theta d\phi\, \cF(p_k+\epsilon e^{i\phi})
\ee
The first term is constant along the integration contour in (\ref{eq:DeltaB}) and does not complicate the integration there.
The second term in (\ref{eq:Iexpl}) can be evaluated explicitly, since the contour is localized around $p_k$ such that the integrand can be expanded.
The contribution to $\Delta_k \cB+\mathrm{c.c.}$ then becomes
\be
 \int_{C_k} dz \, \cI(\eta_-\partial_z\cA_+^s-\eta_+\partial_z\cA_-^s)+\mathrm{c.c.}=
 i\pi Y^k \Big( \cI(p_k+\epsilon)-\overline{\cI(p_k+\epsilon)}\Big)
 \label{eq:DeltaB2}
\ee
Evaluating $\Delta_k\cB+\Delta_k\bar\cB$ using (\ref{eq:DeltaB}) with (\ref{eq:DeltaB1}) and (\ref{eq:DeltaB2}) shows that it exactly reproduces the already existing terms in (\ref{eq:DeltaG0}).
We thus find for the shift of $\cG$ given in (\ref{eq:DeltaG0})
\begin{align}
 \frac{\Delta_k\cG}{2\pi i}&=
 2\bar \cA^0 Y_+^k+2\cA^0 Y_-^k+\sum_{\ell\neq k} Y^{[\ell, k]}\ln |p_\ell-p_k|^2
 +Y^k\left(\overline{\cI(p_k+\epsilon)}-\cI(p_k+\ep)\right)
 \nonumber\\&\hphantom{=}
 +f(p_k) Y^k\left(\eta_-\cA_+^s(p_k+\ep)-\overline{\eta_+\cA_-^s(p_k+\ep)}+\mathrm{c.c.}\right)
 \label{eq:DeltaG0a}
\end{align}
Using that $f(p_k)$ is imaginary, we can write the shift in $\cG$ in the form given in (\ref{eq:DeltaG0b}), completing the derivation for that result.

The second result for which we have not provided a detailed derivation in the main part concerns the integration by parts in (\ref{eq:DeltaG0b}), to arrive at (\ref{eq:DeltaG0c}).
We repeat (\ref{eq:DeltaG0b}) for convenience
\begin{align}
 \frac{\Delta_k\cG}{2\pi i}&=
 2\bar \cA^0 Y_+^k+2\cA^0 Y_-^k+\sum_{\ell\neq k} Y^{[\ell, k]}\ln |p_\ell-p_k|^2
 \nonumber\\&\hphantom{=}
 +Y^k\left(f(p_k)\Big[\eta_-\cA_+^s(p_k+\ep)-\eta_+\cA_-^s(p_k+\ep)\Big]-\cI(p_k+\ep)-\mathrm{c.c.}\right)
 \label{eq:DeltaG0b-app}
\end{align}
The individual terms in the second line are divergent as $\epsilon\rightarrow 0$, but their combination is finite.
We can use $\cA_+^{(0)}-\cA_-^{(0)}=\eta_-\cA_+^s-\eta_+\cA_-^s$
and integration by parts to rewrite $\cI$ defined in (\ref{eq:I}) as
\begin{align}
\label{eq:ibp}
\cI(p_k+\epsilon)&=\int_\infty^{p_k+\epsilon}dw\, f(w) \left(\eta_-\partial_w\cA_+^s-\eta_+\partial_w\cA_-^s\right)
\no\\
&=f(w)\left(\eta_-\cA_+^s-\eta_+\cA_-^s\right)\Big\vert_{\infty}^{p_k+\epsilon}
-\int_\infty^{p_k+\epsilon}dw \left(\eta_-\cA_+^s-\eta_+\cA_-^s\right)\partial_w f
\end{align}
The first term evaluated at $p_k+\epsilon$ cancels the first term in the round brackets in the second line of (\ref{eq:DeltaG0b-app}).
The first term evaluated at $\infty$ becomes $(\eta_-\cA_+^0-\eta_+\cA_-^0)f(+\infty)$, since $\sum_\ell Y_\pm^\ell=0$.
The integrand of the last term in (\ref{eq:ibp}) is only logarithmically divergent as $p_k$ is approached and in particular integrable,
so we can now drop $\epsilon$ in the integration bound.
The shift (\ref{eq:DeltaG0b-app}) therefore becomes
\begin{align}
 \frac{\Delta_k\cG}{2\pi i}&=
 2\bar \cA^0 Y_+^k+2\cA^0 Y_-^k+\sum_{\ell\neq k} Y^{[\ell, k]}\ln |p_\ell-p_k|^2
 \nonumber\\&\hphantom{=}
 +Y^k\left(-(\eta_-\cA_+^0-\eta_+\cA_-^0)f(+\infty)+\int_\infty^{p_k}dw \left(\eta_-\cA_+^s-\eta_+\cA_-^s\right)\partial_w f-\mathrm{c.c.}\right)
\end{align}
The integral in the second line contains the integration constants $\cA_\pm^0$, which only multiply $\partial_w f$,
and it will be convenient to extract them.
This yields
\begin{align}
 \frac{\Delta_k\cG}{2\pi i}&=
 2\bar \cA^0 Y_+^k+2\cA^0 Y_-^k+\sum_{\ell\neq k} Y^{[\ell, k]}\ln |p_\ell-p_k|^2
 \nonumber\\&\hphantom{=}
 +2f(p_k)Y^k(\eta_-\cA_+^0-\eta_+\cA_-^0)
 +Y^k\left(\int_\infty^{p_k}dw \sum_{\ell=1}^L Y^\ell \ln(w-p_\ell) \partial_w f-\mathrm{c.c.}\right)
\end{align}
This is the result quoted in (\ref{eq:DeltaG0c}), thus completing the derivation for that result as well.

\newpage

\bibliographystyle{JHEP.bst}
\bibliography{ads6}
\end{document}